\documentclass{article}
\usepackage[margin=1.25in]{geometry}
\usepackage{graphicx}
\graphicspath{{../BSSS_Ascent_Figs/}{BSSS_Ascent_Figs/}}
\usepackage{amsmath}
\usepackage{amssymb}
\usepackage{mathtools}
\usepackage{parskip}
\usepackage[hidelinks]{hyperref}
\usepackage{natbib}
\usepackage{subcaption}
\usepackage{wrapfig}
\usepackage{color}
\usepackage{amsthm}
\usepackage{wasysym}
\usepackage{ulem}
\usepackage{booktabs}

\emergencystretch=\maxdimen
\begin{document}
\title{On minimizing cyclists' ascent times: Part~I}
\author{
	Len Bos%
	\footnote{
		Universit\`a di Verona, Italy, \texttt{leonardpeter.bos@univr.it}
	}\,,
	Michael A. Slawinski%
	\footnote{
		Memorial University of Newfoundland, Canada, \texttt{mslawins@mac.com}
	}\,,\
	Rapha\"el A. Slawinski%
	\footnote{
		Mount Royal University, Canada, \texttt{rslawinski@mtroyal.ca}
	}\,,\
	Theodore Stanoev%
	\footnote{
		Memorial University of Newfoundland, Canada, \texttt{theodore.stanoev@gmail.com}
	}
}
\date{}
\maketitle
\begin{abstract}
We prove that, given an average power, the ascent time is minimized if a cyclist maintains a constant ground speed regardless of the slope.
Herein, minimizing the time is equivalent to maximizing\,---\,for a given uphill\,---\,the corresponding mean ascent velocity (VAM: {\it velocit\`a ascensionale media\/}), which is a common training metric.
We illustrate the proof with numerical examples, and show that, in general, maintaining a constant instantaneous power results in longer ascent times; both strategies result in the same time if the slope is constant.
To remain within the athlete's capacity, we examine the effect of complementing the average-power constraint with a maximum-power constraint.
Even with this additional constraint, the ascent time is the shortest with a modified constant-speed\,---\,not constant-power\,---\,strategy; as expected, both strategies result in the same time if the maximum and average powers are equal to one another.
Given standard available information\,---\,including level of fitness, quantified by the power output, and ascent profile\,---\,our results allow to formulate reliable and convenient strategies of uphill timetrials.  
\end{abstract}
\section{Introduction}
In this article, we formulate a strategy to minimize the time of climbing an uphill with a bicycle, which is the goal for timetrial competitions composed of an uphill; in Italy, they are referred to with the specific name of a {\it cronoscalata\/}.
We consider that the ascent, including distance and steepness, is known.
 The weight of the cyclist, together with the bicycle, as well as fitness level, quantified by the power output, are also known.
With these\,---\,and other information, such as the air, rolling and drivetrain resistances\,---\,we formulate the aforementioned strategy.
The formulation is based on power considerations, which\,---\,for the last couple of decades\,---\,has  been the key metric in training and race preparation.
Notably, most modern racing bicycles have power meters.

We begin this article by invoking a phenomenological model to estimate the power required to maintain a given speed.
This model is analogous to the one used and discussed by \cite{BosEtAl2023,BosEtAl2024}.
Subsequently, we seek the riding strategy that results in the least ascent time, which\,---\,under stated constraints and assumptions\,---\,we show to be constant speed.
We proceed to exemplify this result for several distinct ascents.
We conclude by discussing the obtained result, comparing it to another plausible strategy: constant power, and commenting on possible future developments.
In the appendices, we provide a method for imposing an optimization constraint, namely, the value of the available average power, 
justify an aspect of our model simplification, namely, neglecting the power used to increase kinetic energy, 
examine the effect of imposing the second constraint, namely, the value of the available maximum power, comment on numerical optimization,  discuss an {\it ad hoc} constant-ascent-speed strategy, give an insight into empirical adequacy of the model, present an analytical Euler-Lagrange solution for optimal ground speed, and conclude with a formulation of an inverse problem to estimate the model parameters by minimizing the discrepancy between model predictions and measurements.
\section{Phenomenological model}
\label{sec:Model}
\begin{figure}
	\centering
	\includegraphics[width=0.43\textwidth]{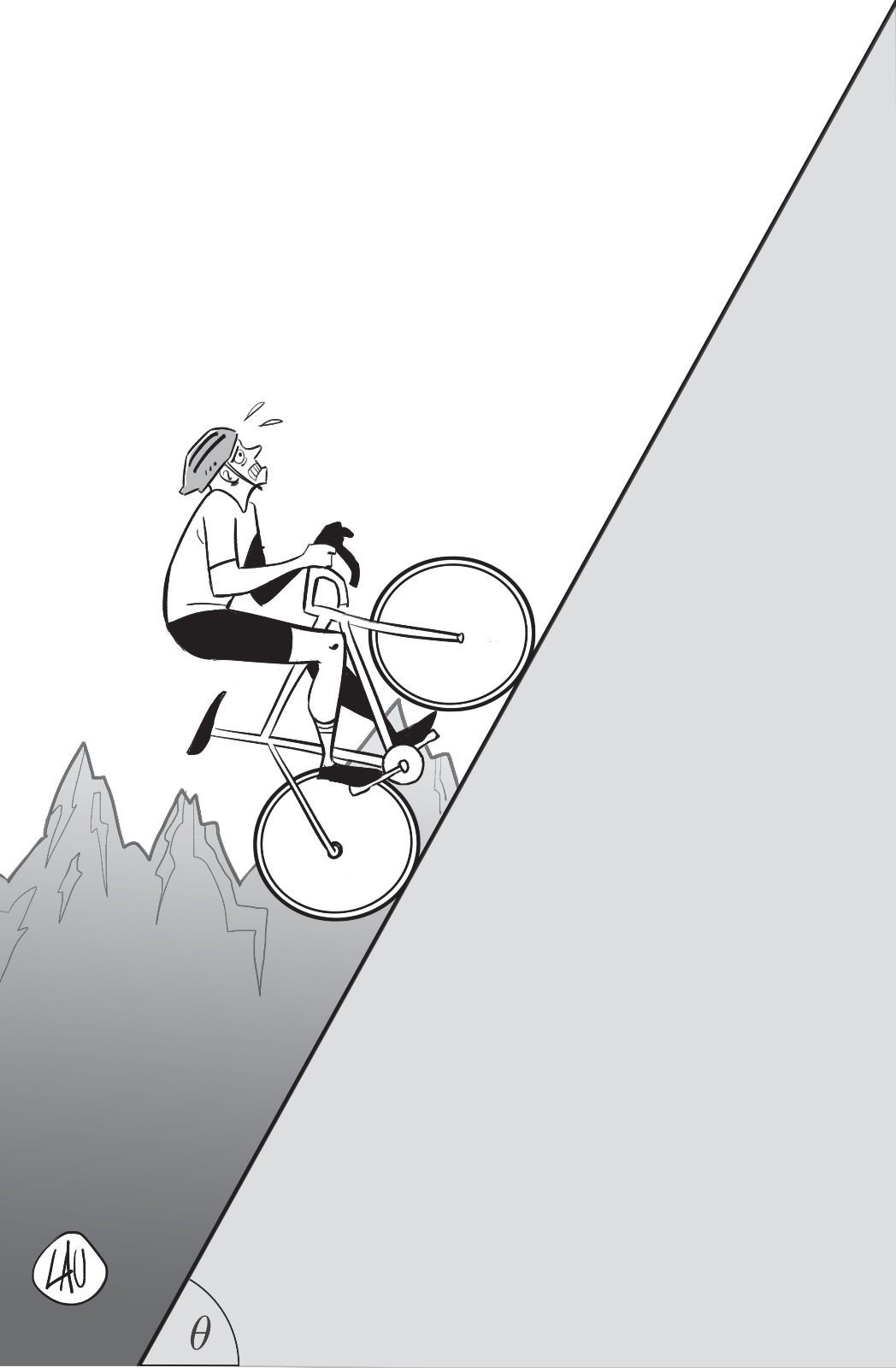}
	\caption{An ascent~$\ldots$}
	\label{fig:Fig_BSSS_Roberto}
\end{figure}

We consider a phenomenological model for a bicycle-cyclist system of mass~$m$ moving with ground speed $V$.%
\footnote{To be consistent with our previous study, \citep{BosEtAl2023,BosEtAl2024}, where we distinguish between the wheel speed,~$v$\,, and the centre-of-mass speed,~$V$\,, herein we use implicitly the latter.
In this model, formally, $v\equiv V$\,; however, conceptually, we consider the motion of the centre of mass.}
The effort expended by the cyclist is quantified by power~$P$\,, which is composed of power~$P_K$\,, required to change the kinetic energy,~$\tfrac{1}{2}mV^2$, and the power to overcome opposing forces,~$P_F$\,, as well as to overcome the drivetrain resistance,
\begin{align*}
	P
	&=\dfrac{
     \overbrace{m\dfrac{{\rm d}V}{{\rm d}t}\,V}^{P_K}
    +
    \overbrace{F\,V}^{P_F}}
    {1-\lambda}\,,
\end{align*}
where $\lambda$ is the drivetrain-resistance coefficient.
$P_F$ is affected by change in elevation, as well as the resistance of rolling and of air.
Hence, explicitly,
\begin{align}
\label{eq:PV}
	P
	&=
	\quad\dfrac{
		\!\!\!\overbrace{\vphantom{\left(V\right)^2}\quad m\,\dfrac{{\rm d}V}{{\rm d}t}\quad}^\text{change in speed}
		+
        \overbrace{\vphantom{\left(V\right)^2}\,m\,g\sin\theta\,}^\text{change in elevation}
		+\
		\overbrace{\vphantom{\left(V\right)^2}
			{\rm C_{rr}}\!\!\!\underbrace{\,m\,g\cos\theta}_\text{normal force}
		}^\text{rolling resistance}
		+
		\overbrace{\vphantom{\left(V\right)^2}
			\,\tfrac{1}{2}\,{\rm C_{d}A}\,\rho\,
					V^{2}\,
		}^\text{air resistance}
	}{
		\underbrace{\quad1-\lambda\quad}_\text{drivetrain efficiency}
	}\,V\,,
\end{align}
where $g$ is the acceleration due to gravity, $\theta$ is the slope, as illustrated in Figure~\ref{fig:Fig_BSSS_Roberto}, and $\rho$ is the air density.
Thus, the three model parameters are the aforementioned $\lambda$, together with the rolling-resistance coefficient,~$\rm C_{rr}$, and the air-resistance coefficient,~$\rm C_dA$.
\section{Shortest ascent time}
\label{sec:Shortest}
\subsection{Formulation}
\label{sub:Formulation}
To determine the optimal strategy for minimizing the ascent time, we allow the instantaneous power to vary within the constraint of a given average power to be specified {\it a priori\/} based on the cyclist's capacities for the expected duration of the effort.
This power can be obtained from the cyclist's power profile~\citep[e.g.,][Figure~2]{LeoEtAl2022}, as discussed in Appendix~\ref{app:PP}.
We assume the ascent to be monotonic, and model the hill by $N$ straight-line segments.
Furthermore, we assume that each segment is traversed at a constant speed and we allow speed to be discontinuous between segments.\footnote{%
The discontinuous speed results in a simplified model that is easier to understand, yet captures the essence of the physics of more sophisticated models.
In particular, we have analyzed models with continuous speed\,---\,for which there does not seem to be an analytical solution\,---\,and have confirmed by numerical experimentation that the conclusions remain substantially unchanged.}
In the limit\,---\,as $N\rightarrow\infty$\,---\,any realistic hill as well as any speed can be approximated with arbitrary accuracy.
Yet, our argument presented in this section is independent of the value of~$N$.

We neglect $P_K$, the power associated with changes in kinetic energy, which is the first term in the numerator of expression~(\ref{eq:PV}).
This is justified by the fact that, riding uphill, speeds are typically low, and changes in kinetic energy are much smaller than the work done against opposing forces, especially against gravity.
An insight into this issue, albeit indirect and {\it a posteriori\/}, is presented in Appendix~\ref{app:KE}.
Given the low speeds, including changes in kinetic energy would be expected to modify the resulting values only slightly, without substantially altering the conclusions.
Incorporating these changes is a subject of future work.

Given the model and assumptions described above, we wish to find\,---\,for each segment\,---\,the speed, with its corresponding power, that minimize the total ascent time, subject to the constraint of a given average power.
In Section~\ref{sec:Optimum}, below, we find the solution that optimizes the time.
In Section~\ref{sec:Minimum}, we show that the optimum solution is indeed a global minimum.
\subsection{Critical point}
\label{sec:Optimum}
Let the $j$th segment, with length~$L_j$ and slope angle~$\theta_j$, be traversed with speed~$V_j$ and corresponding power~$P_j=P_j(\theta_j,V_j)$, where $\theta_j$ is a parameter and $V_j$ is a variable.
Hence, the ascent time, for the $j$th segment, is $t_j=L_j/V_j$, and the total time is
\begin{equation}
    \label{eq:T}
    T = \sum\limits_{j=1}^N t_j = \sum\limits_{j=1}^N \dfrac{L_j}{V_j}\,.
\end{equation}
We wish to minimize $T$, subject to the constraint of a given average power,
\begin{equation}
    \label{eq:Pave}
    \overline{P} = \dfrac{W}{T}\,,
\end{equation}
where the total work is
\begin{equation}
    \label{eq:W}
    W = \sum\limits_{j=1}^N P_jt_j = \sum\limits_{j=1}^N P_j\dfrac{L_j}{V_j}\,.
\end{equation}
Mathematically, the problem is to optimize $T\left(\{V_j\}\right)$ as a function of segment speeds,~$\{V_j\}$, subject to the constraint
\begin{equation}
	\label{eq:PaveP0}
	\overline{P}\left(\{V_j\}\right)-P_0=0\,,
\end{equation}
where $P_0$ is the specified average power.

Such a constrained optimization problem can be solved using the method of Lagrange multipliers; herein, the Lagrangian function is
\begin{equation}
    \label{eq:Lag}
    \mathcal{L}\left(\{V_j\},\Lambda\right) = T+\Lambda\left(\overline{P}-P_0\right),
\end{equation}
where $\Lambda$ is a Lagrange multiplier introduced to enforce the constraint.
A necessary condition for a constrained optimum is that the gradient of $\mathcal{L}$ taken with respect to vector~${\boldsymbol\beta}=[V_1,\ldots,V_N,\Lambda]^T$ be zero,
\begin{equation}
    \label{eq:betaL}
    \nabla_{\!\boldsymbol\beta\,}\mathcal{L}={\bf0}\,;
\end{equation}
which we write explicitly as
\begin{equation*}
	\left\{
    \begin{array}{l}
        \dfrac{\partial\mathcal{L}}{\partial V_j} = 0,\qquad j=1,\ldots,N \\\\
        \dfrac{\partial\mathcal{L}}{\partial\Lambda} = 0
    \end{array}
    \right.
    \quad.
\end{equation*}
Equivalently, condition~(\ref{eq:betaL}) may be expressed in terms of the gradient taken with respect to vector~${\boldsymbol V}=[V_1,\ldots,V_N]^T$,
\begin{equation}
	\label{eq:betav}
	\left\{
    \begin{array}{l}
        \nabla_{\boldsymbol V}\mathcal{L} = 0\\\\
        \overline{P}-P_0 = 0
    \end{array}
    \right.
    \quad.
\end{equation}
Using definition~(\ref{eq:Lag}), condition~(\ref{eq:betav}) may be written as
\begin{align}
    \label{eq:Tgrad}
	\nonumber\nabla_{\boldsymbol V} T &= -\Lambda\nabla_{\boldsymbol V}\left(\overline{P}-P_0\right)\\
    &= -\Lambda\left(\dfrac{1}{T}\nabla_{\boldsymbol V} W-\dfrac{W}{T^2}\nabla_{\boldsymbol V} T\right)\\
    \nonumber &= -\dfrac{\Lambda}{T}\left(\nabla_{\boldsymbol V} W-P_0\nabla_{\boldsymbol V} T\right),
\end{align}
where to obtain the second equality we use definition~(\ref{eq:Pave}), and to obtain the third equality, we use expression~(\ref{eq:PaveP0}).
Using equations~(\ref{eq:T}) and (\ref{eq:W}) for $T$ and $W$, respectively, expressed in terms of the segment speeds,~${V_j}$, we obtain
\begin{equation*}
    \dfrac{\partial T}{\partial V_j} = -\frac{L_j}{V_j^2}, \qquad j=1,\ldots,N
\end{equation*}
and
\begin{equation*}
    \dfrac{\partial W}{\partial V_j} = \dfrac{\partial P_j}{\partial V_j}\frac{L_j}{V_j} - P_j\frac{L_j}{V_j^2} , \qquad j=1,\ldots,N.
\end{equation*}
Substituting into equation~(\ref{eq:Tgrad}) and simplifying, we obtain
\begin{equation}
	\label{eq:Tcond}
    T = \Lambda\left(V_j\dfrac{\partial P_j}{\partial V_j}-P_j+P_0\right), \qquad j=1,\ldots,N.
\end{equation}
Using model (\ref{eq:PV})\,---\,without the first term in the numerator\,---\,and the assumption that the speed for each segment is constant, the power for the $j$th segment may be written as
\begin{equation}
	\label{eq:Pj}
    P_j = \dfrac{m\,g\sin\theta_j\,V_j + m\,g\,{\rm C_{rr}}\cos\theta_j\,V_j + \tfrac{1}{2}{\rm C_d A}\,\rho\,V_j^3}{1-\lambda},\qquad j=1,\ldots,N.
\end{equation}
Using expressions~(\ref{eq:Pj}), we can write conditions~(\ref{eq:Tcond}) as
\begin{equation}
	\label{eq:Tvj}
    T = \Lambda\left(\dfrac{{\rm C_dA}\,\rho}{1-\lambda}V_j^3+P_0\right),\qquad j=1,\ldots,N,
\end{equation}
where the total ascent time, stated in expression~(\ref{eq:T}) is a function of all the segment speeds, $T=T\left(\{V_j\}\right)$.
System~(\ref{eq:Tvj}) may be solved numerically for the segment speeds that optimize the ascent time.
However, an immediate consequence of conditions~(\ref{eq:Tvj}) is that
\begin{equation*}
    T = \Lambda\left(\dfrac{{\rm C_dA}\,\rho}{1-\lambda}V_i^3+P_0\right)=\Lambda\left(\dfrac{{\rm C_dA}\,\rho}{1-\lambda}V_j^3+P_0\right)\,,
\end{equation*}
for any pair of segments $i$ and $j$.
It follows immediately that $V_i=V_j$, for any pair of segments.
In other words, {\it for a given average power, the critical ascent time is achieved with a constant ground speed regardless of the slope angle\/}.
This conclusion is a consequence of the following points.
\begin{enumerate}
    \item The power expended against gravity and rolling resistance, both of which depend on the slope angle of segment $\theta_j$, is linear in the segment speed,~$V_j$.
    \item The coefficients of air resistance, together with the air density,~${\rm C_{d}A}\,\rho$, as well as the drivetrain efficiency,~$1-\lambda$, are segment-independent.
\end{enumerate}
If either of these conditions were not satisfied, the critical speed\,---\,while it could still be found by numerically solving system~(\ref{eq:Tcond})\,---\,would not be constant.
\subsection{Minimum}
\label{sec:Minimum}
System~(\ref{eq:Tvj}) is a necessary condition for a point in the $N$-dimensional space of segment speeds $\{V_j\}$ to be a critical point of Lagrangian~(\ref{eq:Lag}), in other words, to be an extremum of time~(\ref{eq:T}), subject to constraint~(\ref{eq:PaveP0}).
To prove that the solution found in Section~\ref{sec:Optimum} is indeed a constrained minimum, we note that there is only one critical point, which is a consequence of the cubic relationship between power and velocity of model~\eqref{eq:PV}. 
Hence, the minimum we seek is either at this critical point or else is on the boundary of the region of feasible candidates.

The lower limit on a segment speed,~$V_j$, is the requirement that the cyclist not go backwards, $V_j\geqslant 0$.
The upper limit is that value of $V_j$ for which the work done over the $j$th segment, $W_j=P_j\,t_j$, is maximized.
Mathematically, there is no such maximum; yet, in practice, the capacity of a rider for work is finite.
Let us call this maximum work $W_{\rm max}$ and, hence, the resulting subspace of the space of segment speeds is defined by the condition that
\begin{equation}
	\label{eq:Wmax}
    W\left(\{V_j\}\right) \leqslant W_{\rm{max}}\,,
\end{equation}
where the total work is given by expression~(\ref{eq:W}).
The work done over the~$k$th segment is maximized if the speed on that segment,~$V_k$, is a maximum and the speeds on the other segments,~$V_j$, where $j\neq k$, are zero, subject to condition~(\ref{eq:Wmax}).
Hence, on the boundary of the feasibility region, at least one of the speeds is zero.
However, if at least one of the speeds is zero, the total ascent time is infinity.
Since, by hypothesis, the total work on the boundary of the feasibility region is finite, $W=W_{\rm max}<\infty$, the average power on the boundary is zero,
\begin{equation*}
    \overline{P}=\dfrac{W}{T}=0.
\end{equation*}
Therefore the constraint of a given average power cannot be satisfied on the boundary of the feasibility region, and so the minimum we seek cannot lie on the boundary.
It follows that the constant-speed solution is a global minimum of the constrained ascent time.

\subsection{Ascent-time surface}
\label{sec:surf}
\begin{figure}
	\centering
	\includegraphics[width=0.49\textwidth]{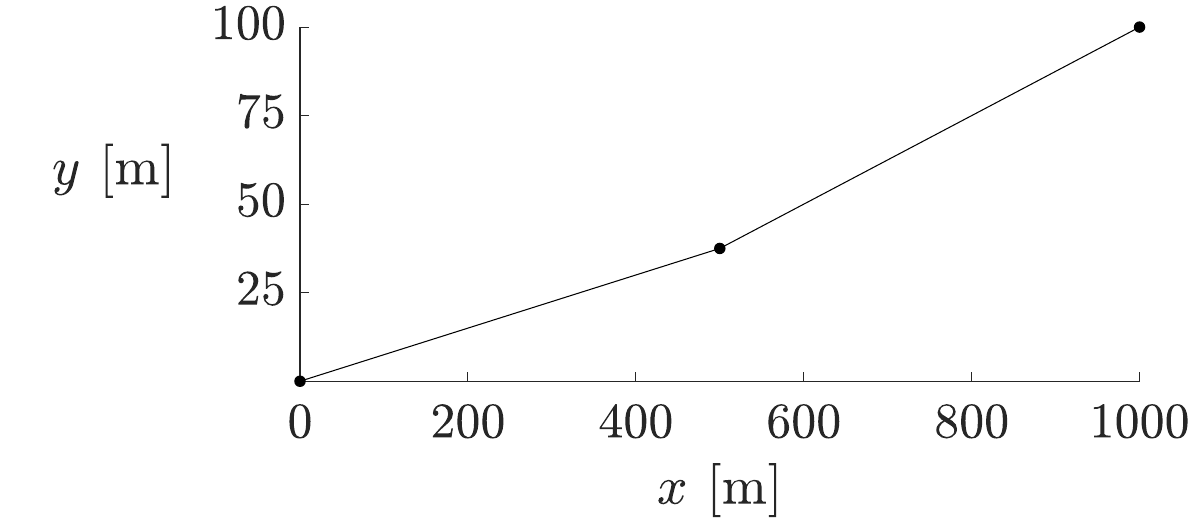}
	\includegraphics[width=0.49\textwidth]{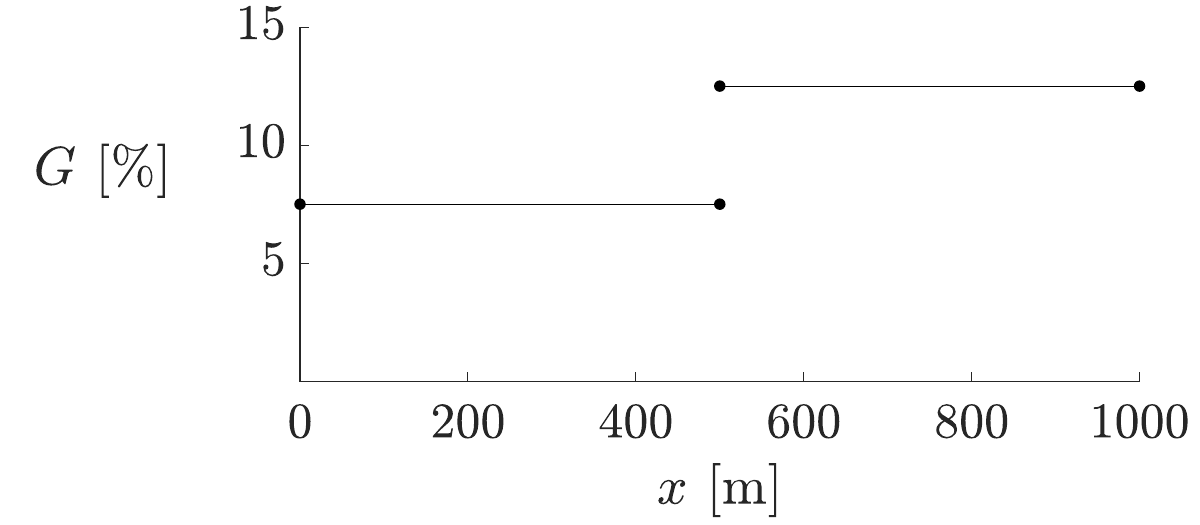}
	\caption{Two-segment ascent and grade}
	\label{fig:twoseg}
\end{figure}

To illustrate the result that a constant speed results in the shortest traveltime, we consider a two-segment ascent,~$N=2$.
We parameterize each segment by its grade, $G_j = \tan\theta_j\cdot100\%$, where $j=1,2$\,, which is the tangent of the slope angle,~$\theta_j$. 
We set the initial coordinates at the origin, $x_0=0\,\rm m$ and $y_0=0\,\rm m$, and the final ones at $x_2=1000\,{\rm m}$ and $y_2=100\,\rm m$.
The intermediate coordinates are $x_1=500\,\rm m$ and $y_1=37.5\,\rm m$, which is tantamount to grade percentages of $G_1=7.5\%$ and $G_2=12.5\%$ and an average of $10\%$.
With these ascent parameters, the length of each segment is $L_1 = 501.4043\,\rm m$ and $L_2 = 503.8911\,\rm m$, which amounts to a total ascent distance of $1005.2954\,\rm m$.
We plot the ascent, and its associated percent grades, in Figure~\ref{fig:twoseg}.

\begin{figure}
	\centering
	\includegraphics[width=0.65\textwidth]{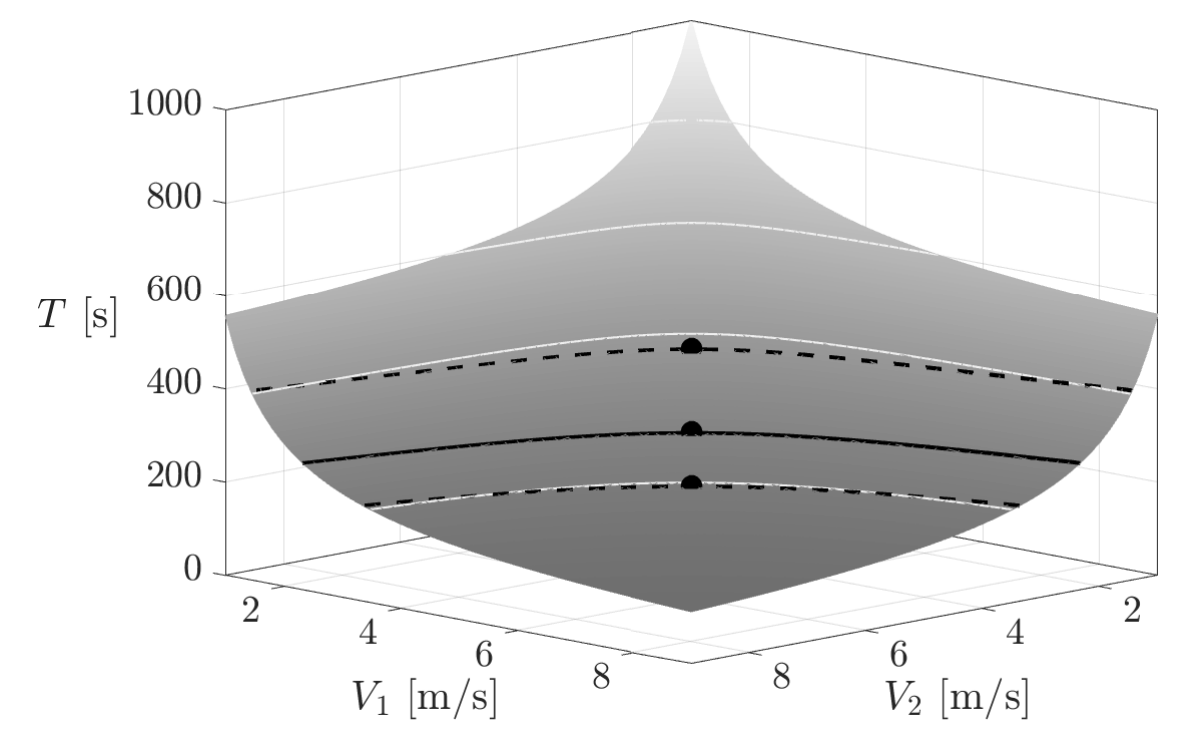}
	\caption{Ascent-time surface: black lines correspond to curves of a given time-averaged power; black dots therein correspond to the $V_1=V_2$ speeds that are solutions of the ascent-time minimization with a constraint of a given power; white lines are level curves of ascent time.}
	\label{fig:surf}
\end{figure}

To calculate the ascent time, stated in expression~\eqref{eq:T}, we let $m=70\,{\rm kg}$, $g = 9.81\,{\rm m/s^2}$, $\rho=1.2\,{\rm kg/m^3}$, ${\rm C_dA} = 0.3\,{\rm m^2}$, ${\rm C_{rr} }= 0.005$ and $\lambda = 0.02$.
In Figure~\ref{fig:surf}, we plot the resulting ascent-time surface as a function of segment speeds, $V_1$ and $V_2$.
Therein, the black curves correspond to three distinct constraints, given by the time-averaged power.
The continuous black line corresponds to $\overline{P}=300\,\rm W$, used in subsequent examples.
The dashed lines correspond to $\overline{P}=200\,\rm W$ and  $\overline{P}=400\,\rm W$.
The level curves of ascent time are white.
The greater the power the shorter the ascent time. 
For each constraint, the minimum ascent times is achieved with~$V_1=V_2$, which correspond to black dots:
for $\overline{P}=200\,\rm W$, $V_1=V_2=2.6842\,\rm m/s$ and $T = 374.5267\,\rm s$;
for $\overline{P}=300\,\rm W$, $V_1=V_2=3.9450\,\rm m/s$ and $T = 254.8288\,\rm s$;
for $\overline{P}=400\,\rm W$, $V_1=V_2=5.1272\,\rm m/s$ and $T = 196.0724\,\rm s$.

\section{Application to arbitrary ascents}
\label{sec:App}
\subsection{Ascent}
\label{sec:ascents}
\begin{figure}
	\centering
	\begin{subfigure}[t]{\textwidth}
		\centering
		\caption{Zero concavity}
		\includegraphics[width=0.49\textwidth]{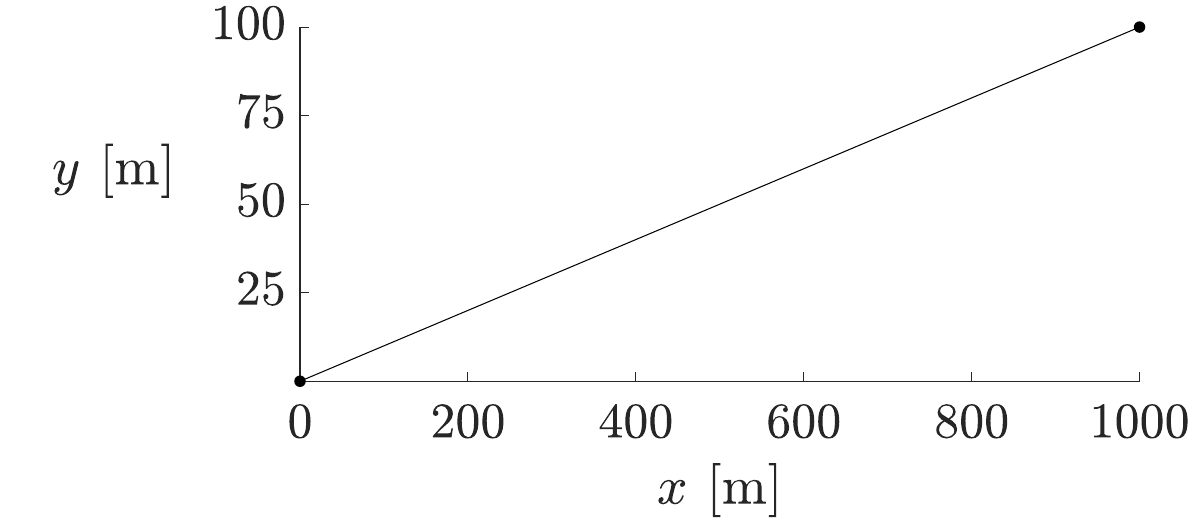}
		\includegraphics[width=0.49\textwidth]{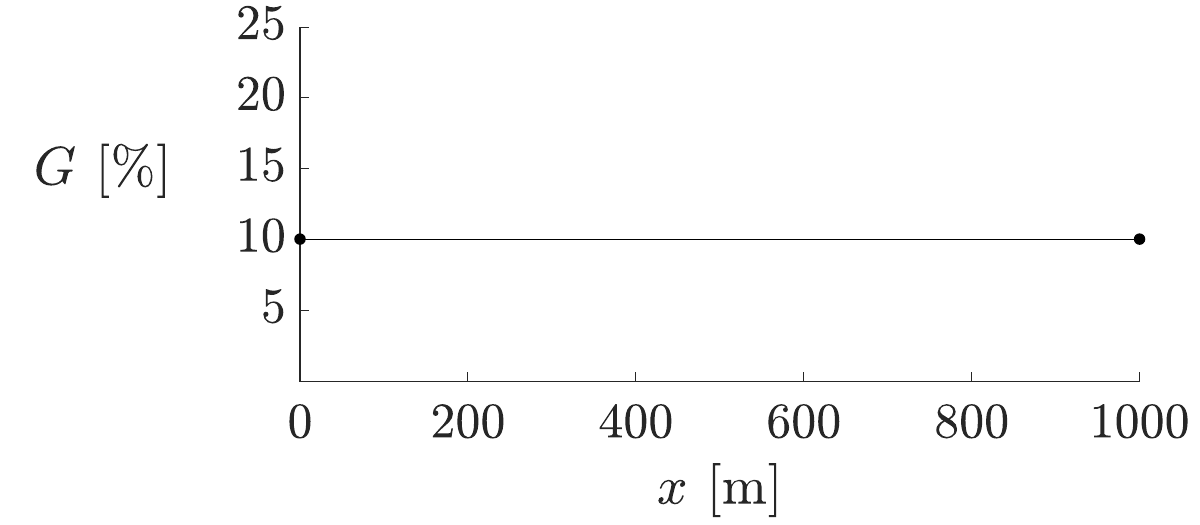}
		\label{fig:ascenta}
	\end{subfigure}
	\begin{subfigure}[t]{\textwidth}
		\centering
		\caption{Positive concavity}
		\includegraphics[width=0.49\textwidth]{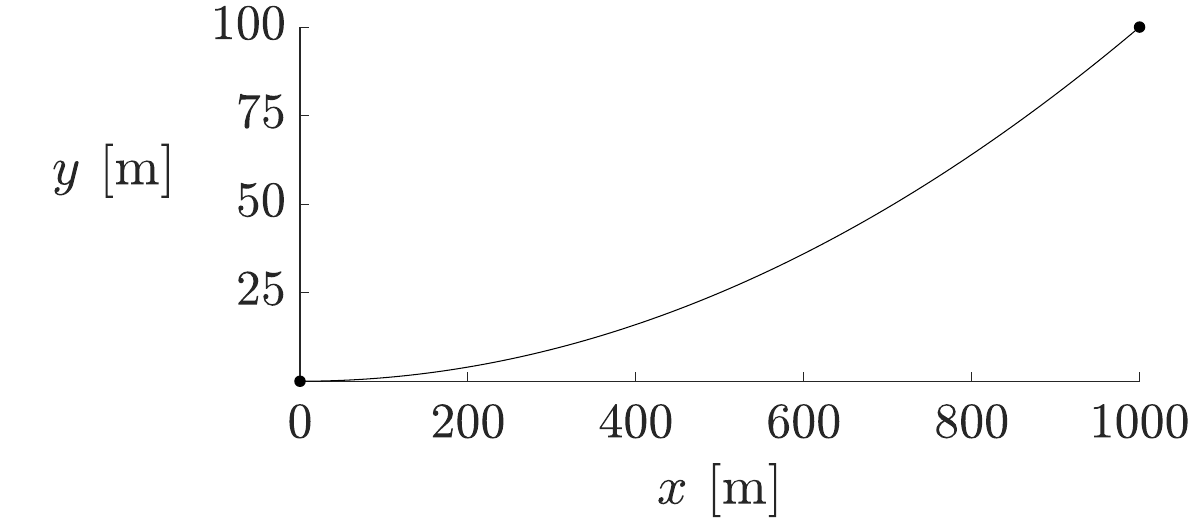}
		\includegraphics[width=0.49\textwidth]{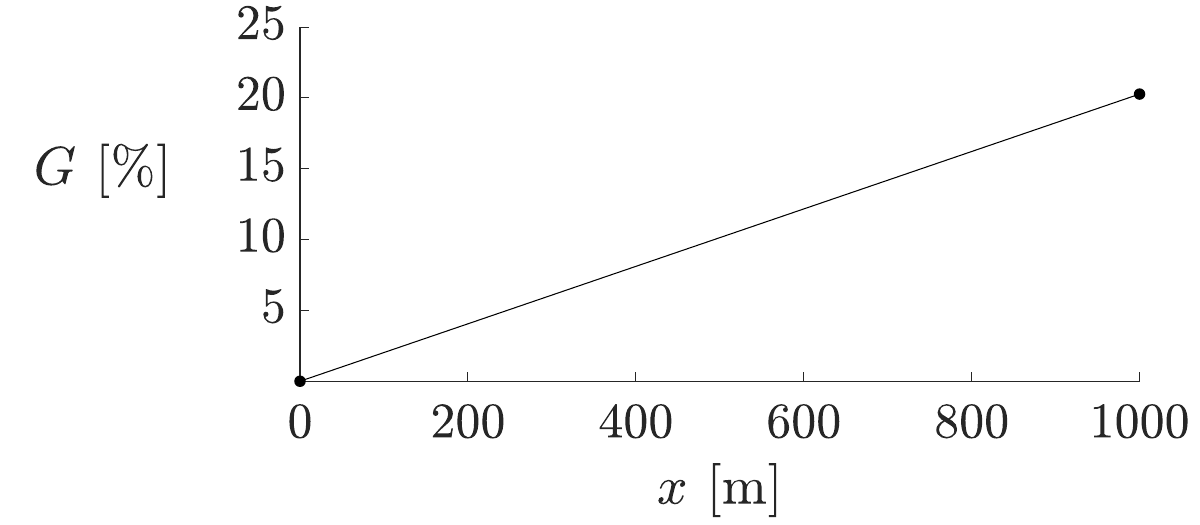}
		\label{fig:ascentb}
	\end{subfigure}
	\begin{subfigure}[t]{\textwidth}
		\centering
		\caption{Negative concavity}
		\includegraphics[width=0.49\textwidth]{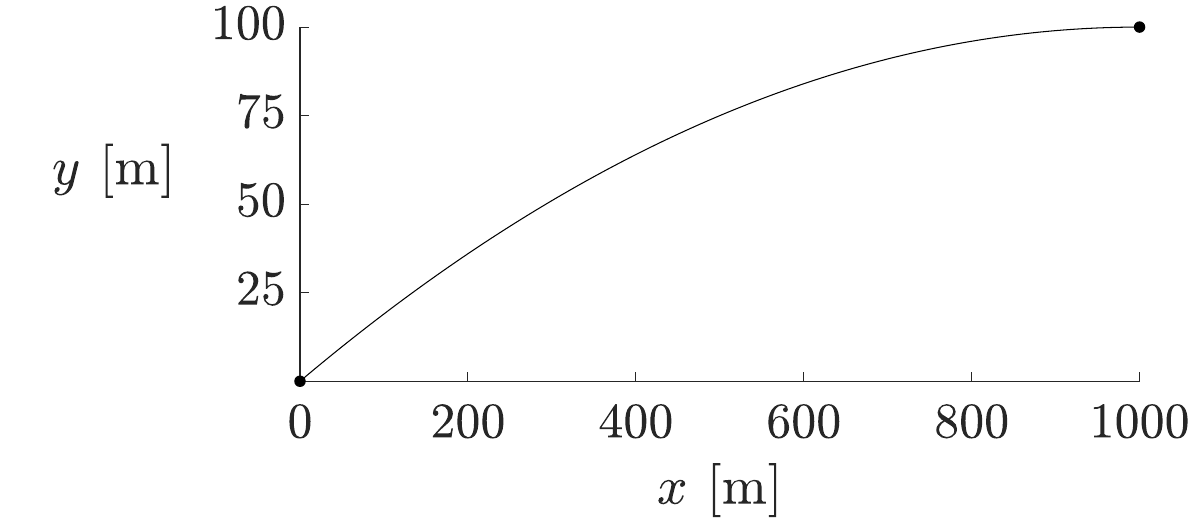}
		\includegraphics[width=0.49\textwidth]{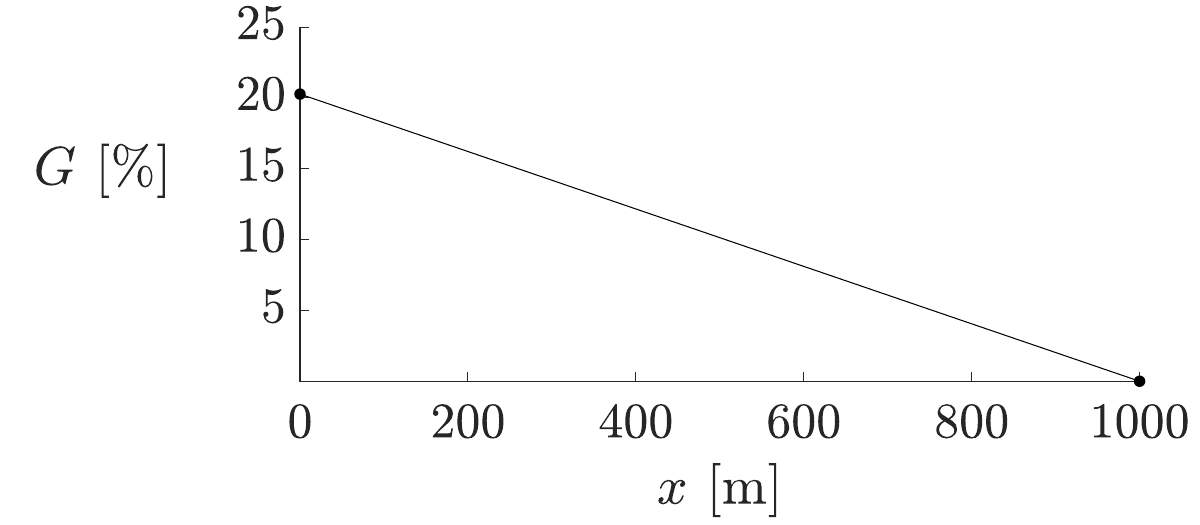}
		\label{fig:ascentc}
	\end{subfigure}
	\begin{subfigure}[t]{\textwidth}
		\centering
		\caption{Positive then negative concavity}
		\includegraphics[width=0.49\textwidth]{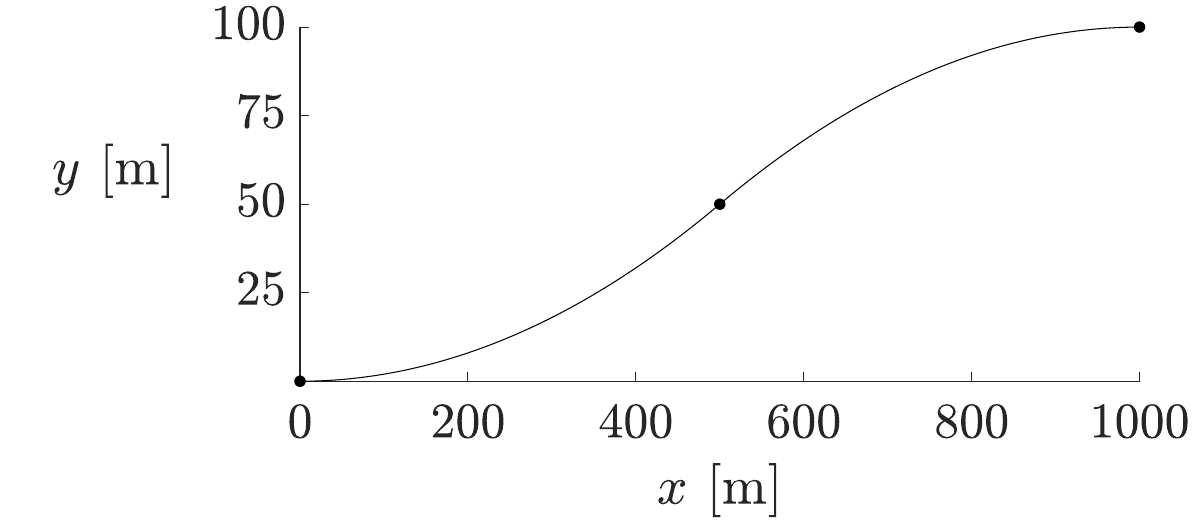}
		\includegraphics[width=0.49\textwidth]{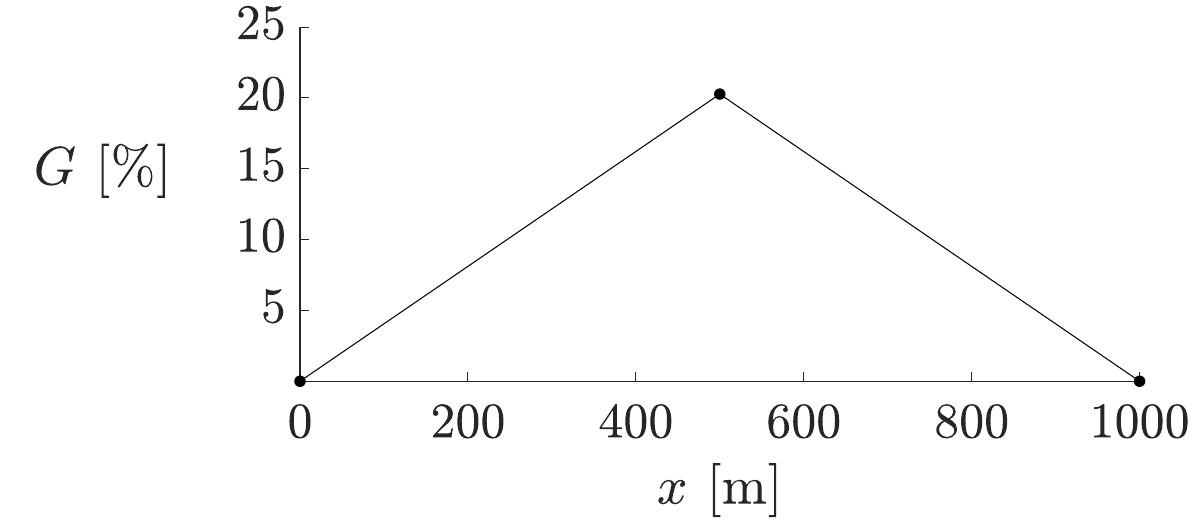}
		\label{fig:ascentd}
	\end{subfigure}
	\begin{subfigure}[t]{\textwidth}
		\centering
		\caption{Negative then positive concavity}
		\includegraphics[width=0.49\textwidth]{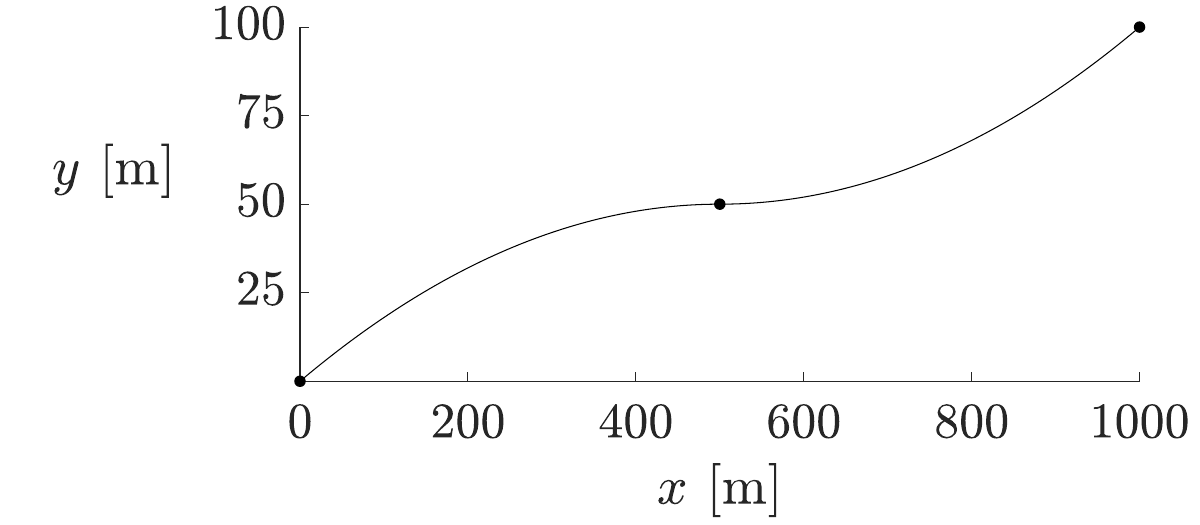}
		\includegraphics[width=0.49\textwidth]{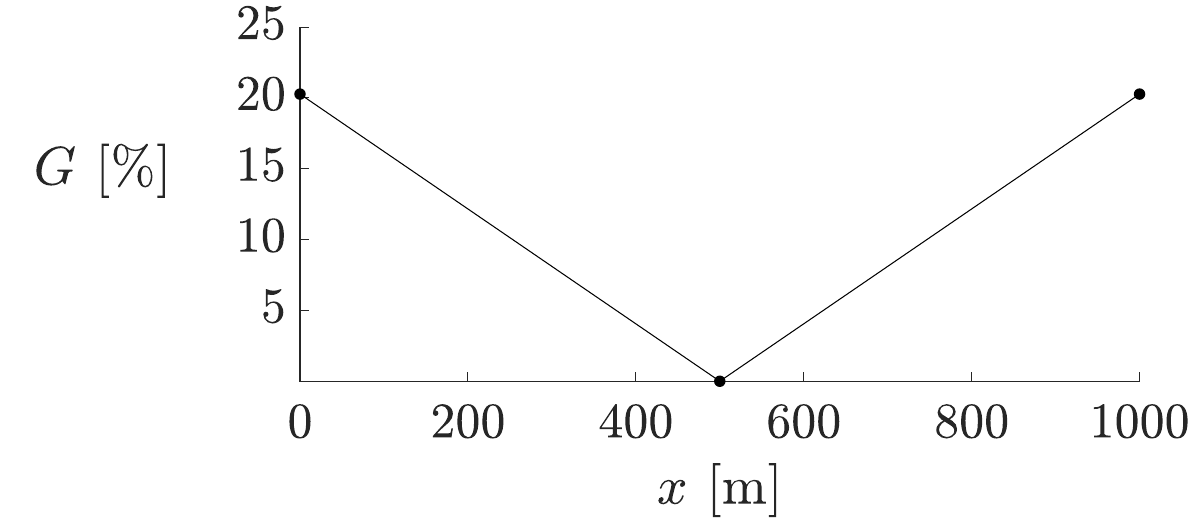}
		\label{fig:ascente}
	\end{subfigure}
	\caption{Ascent and grades of varying concavity}
	\label{fig:ascent}
\end{figure}
As stated in Section~\ref{sec:Shortest}, the method presented therein can be applied to any uphill by discretizing it into $N$ straight-line segments. 
In this section, we apply the method to various ascents.
To compare the optimal strategy of constant speed and variable power with another strategy, we include the ascent times under the assumption of constant power and variable speed.

To generate continuous ascents, which can be sampled with arbitrary resolution, we consider $(n+1)$ nodes in $x\,G$-plane, where $x$ corresponds to the horizontal distance and $G$ is the grade, which is the tangent of the slope angle.
We subdivide the distance with strictly increasing coordinates, $x_0<x_1<\cdots<x_n$. 
Along the $i$th segment, we define the grade as a piecewise linear interpolation of its endpoints,
\begin{equation*}
	 \mathcal{G}_i(X) = G_{i-1} + \left(\dfrac{G_i-G_{i-1}}{x_i-x_{i-1}}\right)(X-x_{i-1}),\qquad
	 x_{i-1}\leqslant X\leqslant x_i\,.
\end{equation*}
Through integration of the grade, we obtain the vertical coordinates of the ascent in the $xy$-plane, where $y$ is the elevation. 
Thus, the change in elevation along the $i$th segment is
\begin{equation*}
	\Delta y_i(X)
	=
	\int\limits_{x_{i-1}}^X\mathcal{G}_i(\xi)\,{\rm d}\xi
	= 
	G_{i-1}(X-x_{i-1})+\dfrac{1}{2}\left(\dfrac{G_i-G_{i-1}}{x_i-x_{i-1}}\right)(X-x_{i-1})^2,\qquad
	x_{i-1}\leqslant X\leqslant x_i\,,
\end{equation*}
which, at the endpoint, is
\begin{equation*}
	\Delta y_i(x_i) 
	= 
	\left(\dfrac{G_{i-1}+G_i}{2}\right)(x_i-x_{i-1})\,.
\end{equation*}
Therefore, the $i$th vertical ascent coordinate is $y_i = y_{i-1} + \Delta y_i(x_i)$ and, given initial coordinates $x_0$ and $y_0$, the final vertical coordinate is
\begin{equation}
	\label{eq:yn_quadratic}
	y_n
	= 
	y_0
	+
	\sum\limits_{i=1}^n\Delta y_i(x_i)\,.
\end{equation}
The average grade percentage for the entire ascent is
\begin{equation*}
	\overline{\mathcal{G}} = \left(\dfrac{y_n-y_0}{x_n-x_0}\right)\cdot100\%
\end{equation*}
and the ascent distance is 
\begin{equation}
	\label{eq:L}
	L 
	=
	\sum\limits_{i=1}^n\left(\,\,\int\limits_{x_{i-1}}^{x_i}\sqrt{1+\mathcal{G}_i(X)^2}\,{\rm d}X\right).
\end{equation}

For the numerical examples, we consider five ascents with a common average grade of $\overline{\mathcal{G}}=10\%$\,, with the same initial and final coordinates as specified in Section~\ref{sec:surf}.
The ascents differ by their concavity, shown in Figure~\ref{fig:ascent}.

The first ascent has zero concavity.
To calculate the required grade, we use $n=1$, set $G_0=G_n$ in expression~\eqref{eq:yn_quadratic} and solve for 
\begin{equation*}
	y_1
	= 
	y_0
	+
	\left(\dfrac{G_0+G_0}{2}\right)(x_1-x_0)
	\implies
	G_0 = \dfrac{y_1-y_0}{x_1-x_0} = 0.1.
\end{equation*}
According to expression~\eqref{eq:L}, the ascent length is 
\begin{equation*}
	L = \int\limits_0^{1000}\sqrt{1+0.1^2}\,{\rm d}x = \sqrt{1.01}\,(1000-0) = 1004.9876\,{\rm m}.
\end{equation*}
The second ascent has positive concavity.
In a similar manner, we use $n=1$, but set $G_0=0$ in expression~\eqref{eq:yn_quadratic}, and solve for $G_1 = 2(y_1-y_0)/(x_1-x_0) = 0.2$.
For the third ascent, which has negative concavity, we set $n=1$ and $G_1=0$, and solve for $G_0=0.2$.
For fourth ascent, which has positive then negative concavity, we set $n=2$, $x_1=(x_2-x_0)/2=500\,{\rm m}$, and $G_0=G_2=0$, and solve for $G_1=0.2$.
For the fifth ascent, which has negative then positive concavity, we set $n=2$, $x_1=500\,{\rm m}$, and $G_1=0$, and solve for $G_0=G_2=0.2$.
Using expression~\eqref{eq:L}, the ascent distance for the latter four ascents is $1006.6272\,{\rm m}$.
\subsection{Speed}
\label{sec:speed}
As demonstrated in Section~\ref{sec:Shortest}, the shortest ascent time is achieved by constant speed.
In this case, constraint~\eqref{eq:PaveP0}\,---\,using time~\eqref{eq:T}, work~\eqref{eq:W} and $V_j = V$ for $j=1,\dots,N$\,---\,is
\begin{equation*}
	0 
	=
	\overline{P}-P_0
	=
	\dfrac{\sum\limits_{j=1}^NP_jL_j}{\sum\limits_{j=1}^NL_j} - P_0\,.
\end{equation*}
Using model~\eqref{eq:PV}, the constraint becomes 
\begin{equation*}
	0 = \dfrac{\sum\limits_{j=1}^N(\alpha_j+\beta V^2)\,V\,L_j}{\sum\limits_{j=1}^NL_j} - P_0,
\end{equation*}
where 
\begin{equation*}
	\alpha_j = \dfrac{m\,g\left({\rm C_{rr}}\cos\theta_j+\sin\theta_j\right)}{1-\lambda}
	\quad\text{and}\quad
	\beta = \dfrac{\tfrac{1}{2}\,{\rm C_dA}\,\rho}{1-\lambda}\,.
\end{equation*}
Through algebraic manipulation, we write the constraint as a depressed cubic equation
\begin{equation}
\label{eq:DepCub}
	0 = V^3 + p\,V + q\,,
\end{equation}
where
\begin{equation}
	\label{eq:pq_vconst}
	p = \dfrac{\sum\limits_{j=1}^N\alpha_jL_j}{\beta\sum\limits_{j=1}^NL_j}
	\quad\text{and}\quad
	q = -\dfrac{P_0}{\beta}.
\end{equation}
Using Cardano's formula~\citep[e.g.,][pp.~112--113]{Tanton2005}, the solution of the cubic is
\begin{equation}
	\label{eq:vpq}
	V 
	= 
	\sqrt[3]{-\dfrac{q}{2}-\sqrt{\left(\dfrac{q}{2}\right)^2+\left(\dfrac{p}{3}\right)^3}} + \sqrt[3]{-\dfrac{q}{2}+\sqrt{\left(\dfrac{q}{2}\right)^2+\left(\dfrac{p}{3}\right)^3}}.
\end{equation}
Since $p>0$ for $\theta_j\in[\,0,\pi/2\,)$, the discriminant of the cubic is strictly positive, which means that the equation has one real and two complex roots.
In other words, the constraint is the intersection of a cubic monomial and a line of negative slope $V^3 = - p\,V - q$, which only has one real solution.
Moreover, since the vertical intercept of the line is $-q = P_0/\beta > 0,$ the real solution is positive, consistent with its physical interpretation as a speed.
\begin{table}[h]
	\centering
	\begin{tabular}{c*{7}{c}}
		& \multicolumn{3}{c}{Constant-speed case} & & \multicolumn{3}{c}{Constant-power case ($P_j=300\,{\rm W}$)} \\
		\cmidrule{2-4}\cmidrule{6-8}
		Ascent & 
		$T\,\,[{\rm s}]$ & $V\,\,[{\rm m/s}]$ & $\max\{P_j\}\,\,[{\rm W}]$ & & 
		$T\,\,[{\rm s}]$ & $\overline{V}_j\,\,[{\rm m/s}]$ & $\max\{V_j\}\,\,[{\rm m/s}]$ \\ 
		\toprule
		(\subref{fig:ascenta}) & $254.8206$ & $3.9439$ & $300.0000$ & & $254.8206$ & $3.9439$ & $3.9439$ \\
		(\subref{fig:ascentb}) & $254.8642$ & $3.9497$ & $567.6278$ & & $263.7539$ & $3.8165$ & $11.2361$ \\
		(\subref{fig:ascentc}) & $254.8642$ & $3.9497$ & $567.6278$ & & $263.7539$ & $3.8165$ & $11.2361$ \\
		(\subref{fig:ascentd}) & $254.8642$ & $3.9497$ & $567.6018$ & & $263.7539$ & $3.8165$ & $11.2351$ \\
		(\subref{fig:ascente}) & $254.8642$ & $3.9497$ & $567.6018$ & & $263.7539$ & $3.8165$ & $11.2351$ \\
		\bottomrule
	\end{tabular}
	\caption{Ascent times for five ascents}
	\label{tab:ascents}
\end{table}

In a similar manner, the constant-power case requires $P_j-P_0 = 0$ for $j=1,\dots,N$. 
Using model~\eqref{eq:PV}, the constraint on the $j$th segment is also a depressed cubic equation, 
\begin{equation}
	\label{eq:Vj_constP}
	0 = V_j^3 + \dfrac{\alpha_j}{\beta}V_j - \dfrac{P_0}{\beta}. 
\end{equation}
Therefore, we calculate the $j$th speed in the constant-power case with expression~\eqref{eq:vpq}, where we replace $p$ with $p_j=\alpha_j/\beta$.
\subsection{Time}
\label{sec:times}
In this section, to calculate the ascent times, we consider the bicycle-cyclist system with the parameter values as specified in Section~\ref{sec:surf}.
We specify the time-average power as $\overline P=300\,{\rm W}$.
\begin{figure}
	\centering
	\begin{subfigure}[t]{\textwidth}
		\centering
		\caption{}
		\includegraphics[width=0.49\textwidth]{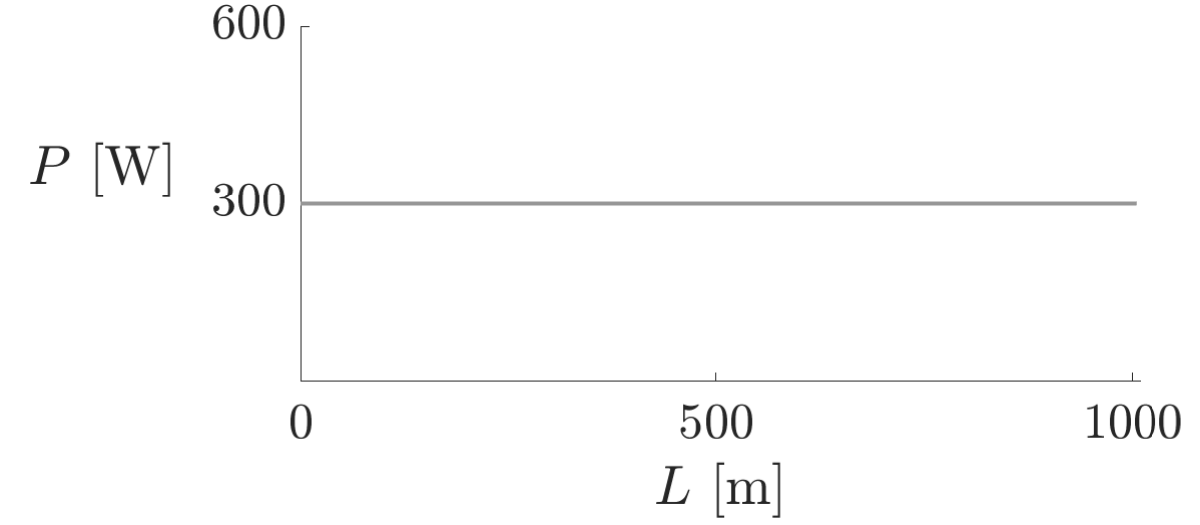}
		\includegraphics[width=0.49\textwidth]{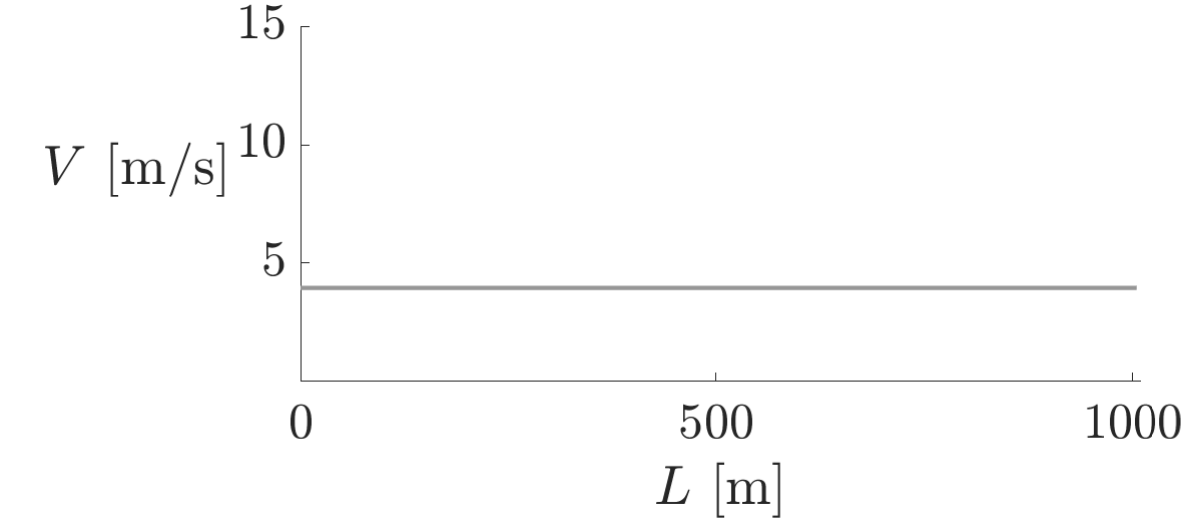}
		\label{fig:PVa}
	\end{subfigure}
	\begin{subfigure}[t]{\textwidth}
		\centering
		\caption{}
		\includegraphics[width=0.49\textwidth]{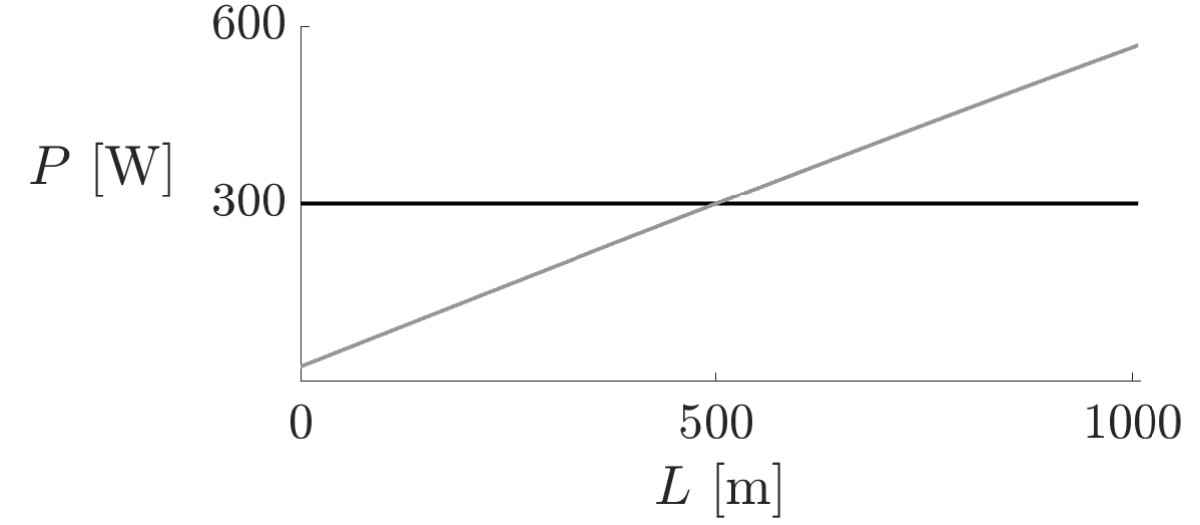}
		\includegraphics[width=0.49\textwidth]{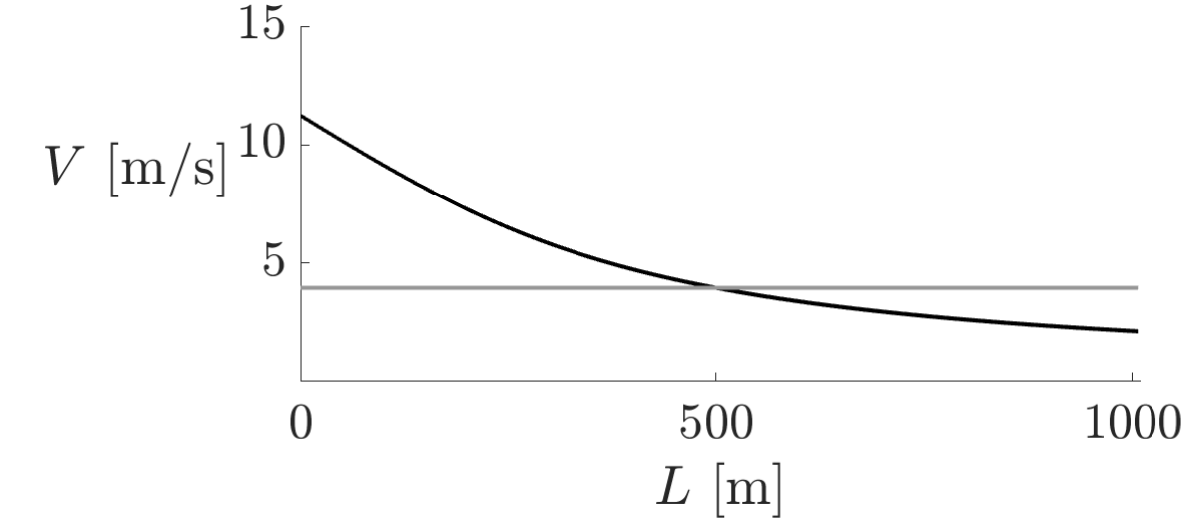}
		\label{fig:PVb}
	\end{subfigure}
	\begin{subfigure}[t]{\textwidth}
		\centering
		\caption{}
		\includegraphics[width=0.49\textwidth]{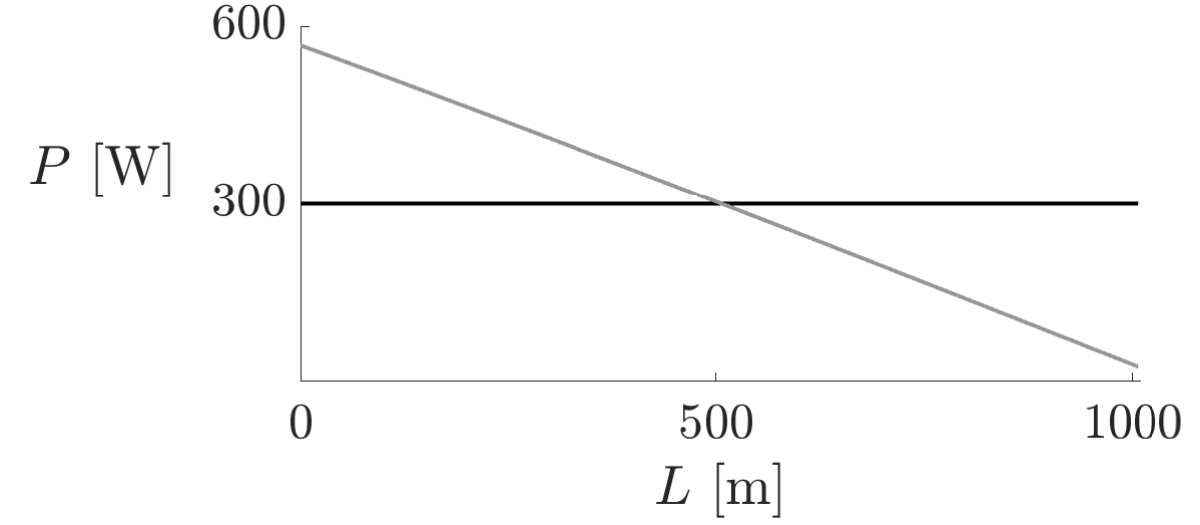}
		\includegraphics[width=0.49\textwidth]{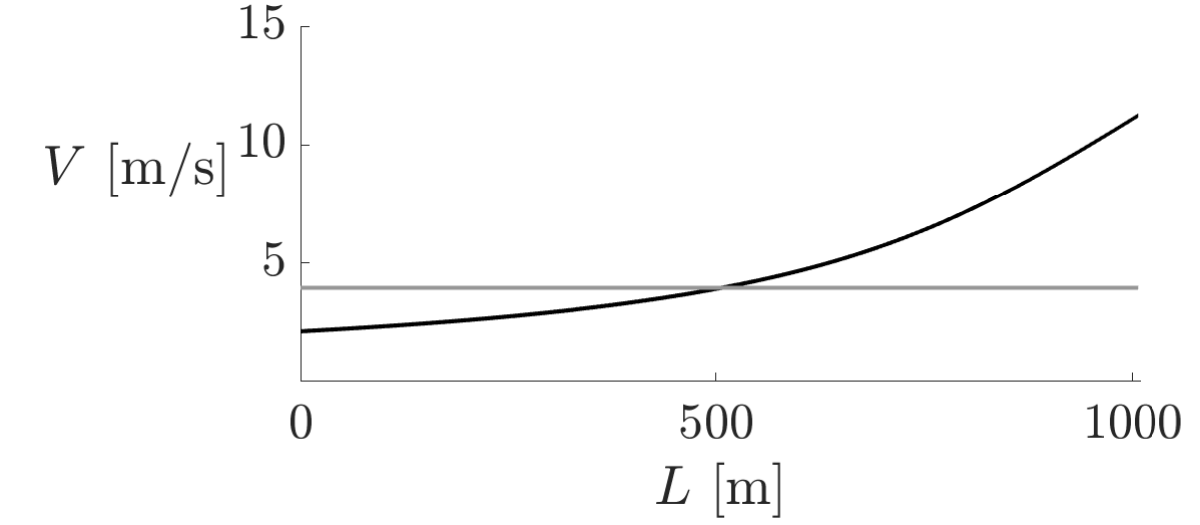}
		\label{fig:PVc}
	\end{subfigure}
	\begin{subfigure}[t]{\textwidth}
		\centering
		\caption{}
		\includegraphics[width=0.49\textwidth]{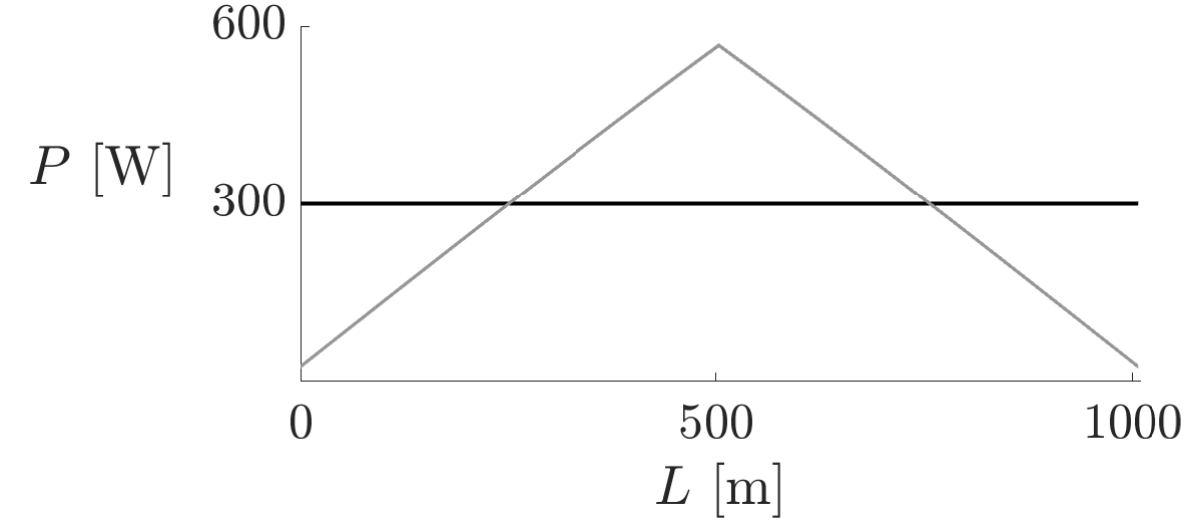}
		\includegraphics[width=0.49\textwidth]{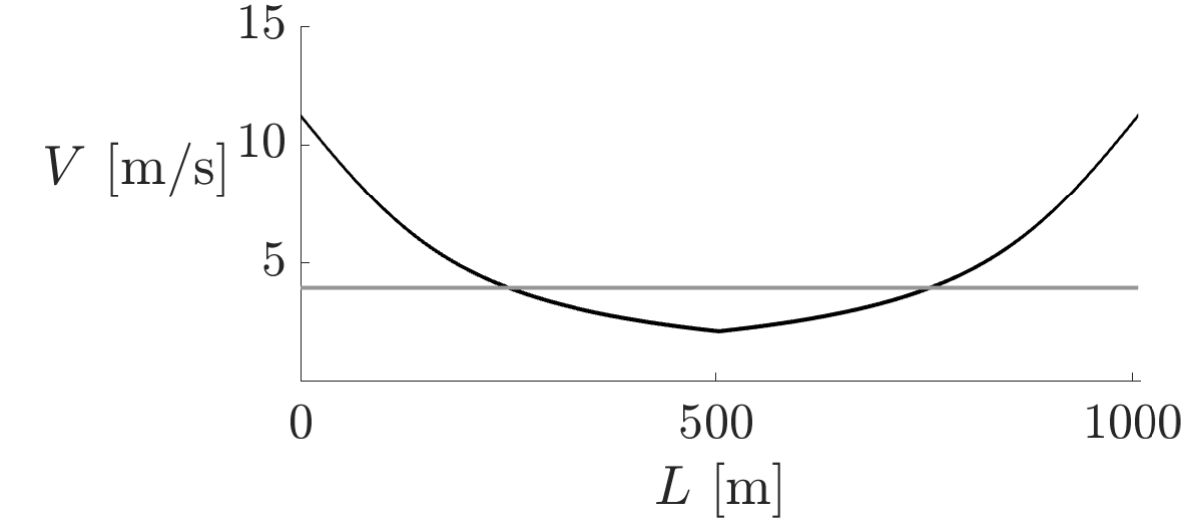}
		\label{fig:PVd}
	\end{subfigure}
	\begin{subfigure}[t]{\textwidth}
		\centering
		\caption{}
		\includegraphics[width=0.49\textwidth]{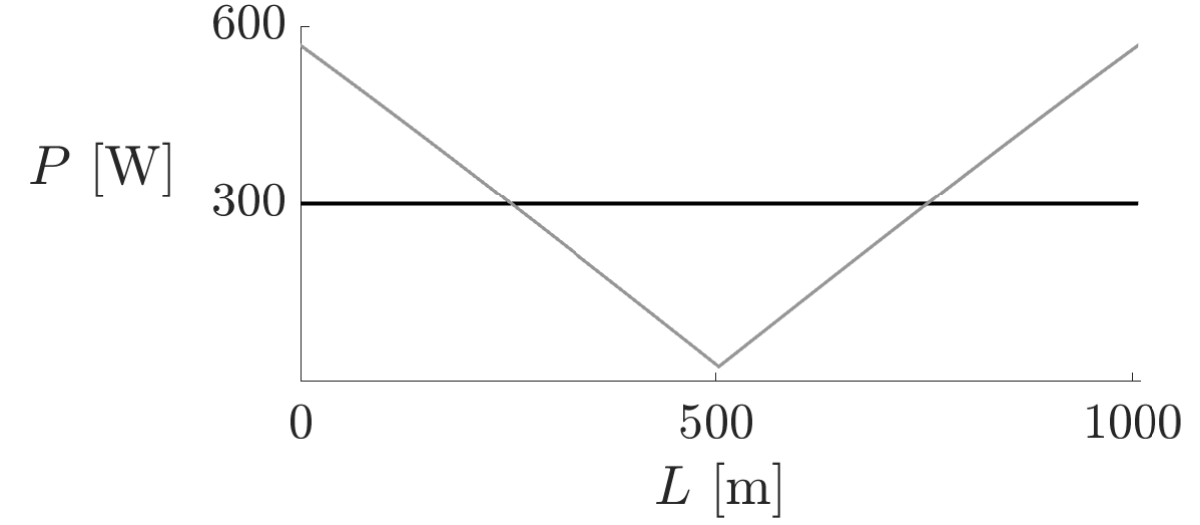}
		\includegraphics[width=0.49\textwidth]{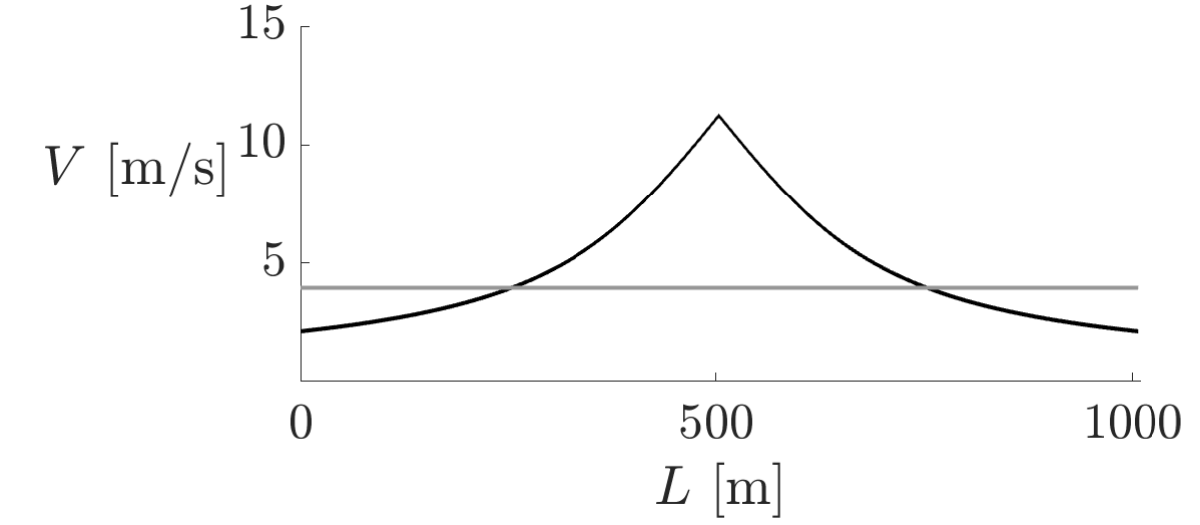}
		\label{fig:PVe}
	\end{subfigure}
	\caption{Power and speed corresponding to the five ascents in Figures~(\ref{fig:ascenta})\,--\,(\ref{fig:ascente}).
	Black lines correspond to the constant-power solution and grey lines to the constant-speed solution; for ascent~(\ref{fig:ascenta}), both have the same solution.
	In contrast to Figure~\ref{fig:ascent}, the horizontal axis is the distance travelled, not its horizontal component.}
\label{fig:PwrSpd}
\end{figure}

For each ascent, we discretize the ascent coordinates using $N=10\,000$ straight-line segments. 
For the constant-speed case, we calculate the speed,~$V$, using expression~\eqref{eq:vpq} with constant-speed~$p$ and~$q$ coefficients of expressions~\eqref{eq:pq_vconst}.%
\footnote{$V$ can be also obtained analytically, as shown in Appendix~\ref{app:E-L}.}
For the constant-power case, we calculate the speed on the $j$th segment,~$V_j$ using expression~\eqref{eq:vpq} with~$p_j = \alpha_j/\beta$ in place of~$p$.

In both cases, we calculate the ascent time,~$T$, using expression~\eqref{eq:T}%
\footnote{For the constant-speed case, $T$ can be also obtained analytically, as shown in Appendix~\ref{app:E-L}.} and tabulate the results in Table~\ref{tab:ascents}.
$T$ is the shortest for ascent~(\subref{fig:ascenta}) because the distance travelled is the shortest.
$T$ is the same for ascents~(\subref{fig:ascentb})--(\subref{fig:ascente}) because their distances are equal, as stated in Section~\ref{sec:ascents}.
In Figure~\ref{fig:PwrSpd}, we present the power and speed for the constant-speed and constant-power cases for ascents~(\subref{fig:ascenta})--(\subref{fig:ascente}).
\section{Discussion and conclusions}
The ascent time is the shortest for the constant-speed case.
For a constant slope, shown in Figure~\ref{fig:ascenta}, constant speed is tantamount to constant power; hence, as shown in the first row of Table~\ref{tab:ascents}, the ascent times are equal to one another for both cases.
For the remaining rows of this table, the constant-speed case results in a time shorter by about~$3.5\%$\,.

The difference between the two strategies increases with the steepness of a portion of an ascent.
Let us consider a three-segment ascent whose overall grade is $10\,\%$, with individual segments of  $2.5\,\%$\,, $25\,\%$ and $2.5\,\%$\,.
Letting $\overline{P}=300\,\rm W$, for a constant-speed strategy, we get $V=3.9631\,\rm m/s$ and $T=254.9668\,\rm s$, with power of individual segments of $94.7177\,\rm W$, $698.4306\,\rm W$ and $94.7177\,\rm W$.
For a constant-power strategy, $T=276.3511\,\rm s$, with speed of individual segments of $8.6395\,\rm m/s$, $1.7252\,\rm m/s$ and $8.6395\,\rm m/s$; since the speed is the same at the bottom and on the top, there is no change in kinetic energy.
In this example, the constant-power strategy is slower by about~$8\%$\,.
For either strategy, variations between the lowest and highest power or speed might be exceedingly large for pragmatic considerations but this example allows us to compare the two strategies and to gain an insight into adjusting them for a given course. 

Commonly, the two strategies\,---\,albeit distinct in requirements imposed upon a rider\,---\,result in similar ascent times, which might allow one to choose the approach preferable for particular circumstances.
Examining similarities of time, however, we need to comment on the assumption discussed in Appendix~\ref{app:KE}.
Not including, in a given constant power, a portion expended to increase kinetic energy\,---\,as is the case for ascent~(\ref{fig:ascentc})\,---\,ignores the fact that this portion is not available for speed.
Hence, the resulting constant-power ascent time is an underestimate; without this assumption, the difference between the two strategies is expected to be greater.
The opposite is true for ascent~(\ref{fig:ascentb}), where the constant-power ascent time is an overestimate.
Including a consideration of kinetic energy is a part of our subsequent work.

Minimizing the ascent time is equivalent to maximizing\,---\,for a given uphill\,---\,the corresponding mean ascent velocity (VAM: {\it velocit\`a ascensionale media\/}), which has typically units of vertical metres per hour. 
Our approach is valid, in general, given an arbitrary uphill.
For a VAM problem, however, we would also look for a specific ascent to maximize the speed of gaining altitude, as discussed by~\citet{BosEtAl2021}.
\begin{figure}
	\centering
	\includegraphics[width=0.9\textwidth]{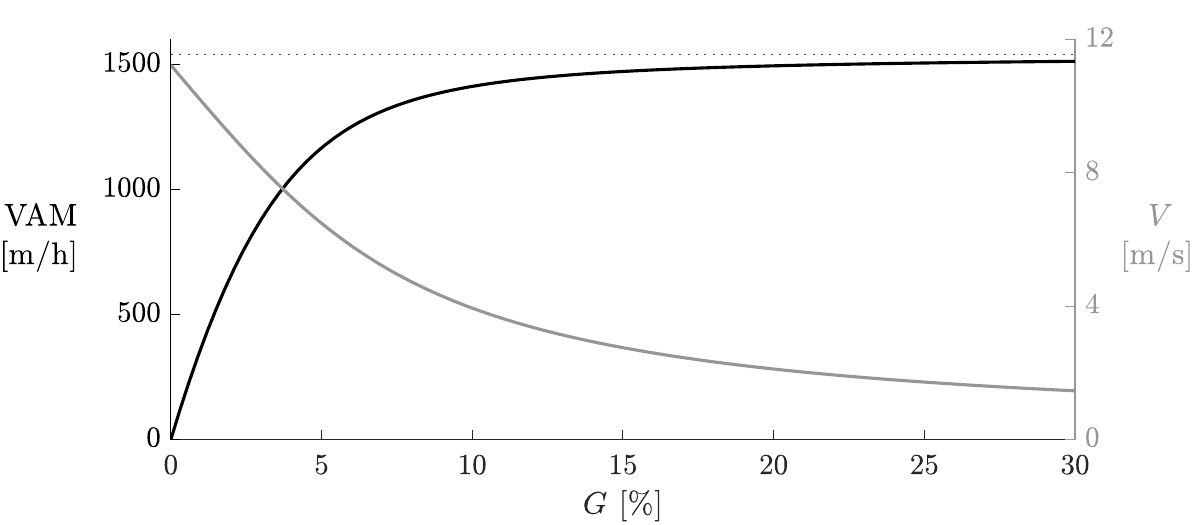}
	\caption{VAM and corresponding ground speed as a function of average power}
	\label{fig:Fig_BSSS_Ascent_VAM}
\end{figure}

Herein we gain certain insights into the VAM problem.
We consider a straight-line ascent with a vertical gain of 100 metres akin to ascent~(\subref{fig:ascenta}), but as a function of grade percentage such that $0 < G < 30\,\%$. 
As shown in Figure~\ref{fig:Fig_BSSS_Ascent_VAM}, under the constraint of a given average power,~$\overline{P}$, there is an upper bound for the vertical speed,~$\rm VAM$, since\,---\,as the grade increases and the speed decreases\,---\,the expended power goes into climbing and overcoming the air and drivetrain resistances, not to doing work against rolling resistance.
In accordance with expression~(\ref{eq:PV}), for $G\to\infty$ and, hence, $\theta\to\pi/2$, 
\begin{equation*}
	P=\dfrac{\,m\,g\,\frac{\rm VAM}{3600} + \tfrac{1}{2}\,{\rm C_{d}A}\,\rho\,V^3\,}{\quad1-\lambda\quad}\,,
\end{equation*}
where
\begin{equation}
	\label{eq:VAM}
	{\rm VAM} := \frac{3600\,\rm s}{\rm h}\,V\sin\theta.
\end{equation}
Hence, with $\lambda=0.02$, $m=70\,{\rm kg}$, $g = 9.81\,{\rm m/s^2}$ and $\overline{P}=300\,\rm W$,
\begin{equation*}
	{\rm VAM}=3600\,\dfrac{(1-\lambda)\,\overline{P}-\tfrac{1}{2}\,{\rm C_{d}A}\,\rho\,V^3}{\,m\,g\,}
	=1541.2104\,\rm m/h\,,
\end{equation*}
which is the sought upper bound, shown as the dotted line in Figure~\ref{fig:Fig_BSSS_Ascent_VAM}.

Furthermore, as shown in Figure~\ref{fig:Fig_BSSS_Ascent_VAM}\,---\,for a given power\,---\,VAM increases monotonically with the grade, which is consistent with~\citet[Theorem~1]{BosEtAl2021}.
The increase is not linear: VAM increases quickly for $0<G<10\%$\,, but the gains in VAM are marginal for $G>10\%$\,.
Also, as shown in Table~\ref{tab:ascents}, the shortest time corresponds to ascent~(\ref{fig:ascenta}), which is a straight line; again, this is consistent with~\citet[Theorem~2]{BosEtAl2021}.

For the proof presented in Section~\ref{sec:Shortest}, with its results illustrated in Section~\ref{sec:App}, the only constraint of the sought minimum is the average power.
Within this sole constraint, the optimal constant speed might require excessive power on steeper portions of the ascent.
As shown on the left-hand plots of Figures~(\ref{fig:PVb})\,--\,(\ref{fig:PVe}), the maximum power reaches almost $570\,\rm W$, as stated in Table~\ref{tab:ascents}, which might be more than a rider can generate given that $\overline{P}=300\,\rm W$.
The effect of including an upper power limit as another constraint is examined in Appendix~\ref{app:Pmax}, and its numerical implementation is discussed in Appendix~\ref{app:NumCon}.

In Appendix~\ref{app:VertSpeed}, we consider an {\it ad hoc} strategy, which, in contrast to constant~$V$, does not emerge from a solution of an optimization problem; its inclusion is motivated by questions from our readers.
That strategy requires a constant ascent speed, which is the vertical component of~$V$.
As we point out in the appendix, that strategy breaks down for ascents with horizontal portions.
Furthermore\,---\,even for ascents without such portions\,---\,it is much inferior to either a constant-speed or constant-power strategy.

Following recent measurements, in Appendix~\ref{app:EmpAdeq}, we examine briefly the empirical adequacy of our model.
We do so for the constant-power strategy, which is easier to keep for a rider since it corresponds to steady effort.
\section*{Acknowledgements}
We wish to acknowledge Matteo Bertrand and Alberto Demicheli, guided by G.\,Andrea Oliveri, for providing the measurements of the ascents, David Dalton, for his proofreading, Elena Patarini, for her graphic support, and Roberto Lauciello, for his artistic contribution.
We also wish to acknowledge insightful comments of an anonymous reviewer of~\citet{BosEtAl2024_DRNA} that led to the derivation presented in Appendix~\ref{app:EL}.
\bibliographystyle{apa}
\bibliography{BSSS_Ascent_arXiv.bib}

\begin{thebibliography}{}

\bibitem[\protect\astroncite{Bos et~al.}{2023}]{BosEtAl2023}
Bos, L., Slawinski, M.~A., Slawinski, R.~A., and Stanoev, T. (2023).
\newblock Modelling of a cyclist's power for time trials on a velodrome.
\newblock {\em ar{X}iv}, 2201.06788 [physics.class-ph].

\bibitem[\protect\astroncite{Bos et~al.}{2024a}]{BosEtAl2024}
Bos, L., Slawinski, M.~A., Slawinski, R.~A., and Stanoev, T. (2024a).
\newblock Modelling of a cyclist's power for time trials on a velodrome.
\newblock {\em Sports Engineering}, 27(9).

\bibitem[\protect\astroncite{Bos et~al.}{2024b}]{BosEtAl2024_DRNA}
Bos, L., Slawinski, M.~A., Slawinski, R.~A., and Stanoev, T. (2024b).
\newblock On minimizing cyclists' ascent times: Part~{I}.
\newblock {\em Dolomites Research Notes on Approximation}, 17(3):5--19.

\bibitem[\protect\astroncite{Bos et~al.}{2021}]{BosEtAl2021}
Bos, L., Slawinski, M.~A., and Stanoev, T. (2021).
\newblock On maximizing {VAM} for a given power: {S}lope, cadence, force and
  gear-ratio considerations.
\newblock {\em ar{X}iv}, 2006.15816 [physics.class-ph].

\bibitem[\protect\astroncite{Danek et~al.}{2021}]{DanekEtAl2021}
Danek, T., Slawinski, M.~A., and Stanoev, T. (2021).
\newblock On modelling bicycle power-meter measurements.
\newblock {\em ar{X}iv}, 2103.09806 [physics.pop-ph].

\bibitem[\protect\astroncite{Leo et~al.}{2022}]{LeoEtAl2022}
Leo, P., Spragg, J., Podlogar, T., Lawley, J.~S., and Mujika, I. (2022).
\newblock Power profiling and the power-duration relationship in cycling: a
  narrative review.
\newblock {\em European Journal of Applied Physiology}, 122:301--316.

\bibitem[\protect\astroncite{MATLAB}{2022}]{MATLAB}
MATLAB (2022).
\newblock {\em version 9.12.0.1884302 (R2022a)}.
\newblock The MathWorks Inc.

\bibitem[\protect\astroncite{Tanton}{2005}]{Tanton2005}
Tanton, J.~S. (2005).
\newblock {\em Encyclopedia of Mathematics}.
\newblock Facts On File, Inc.

\end{thebibliography}
\begin{appendix}
\setcounter{figure}{0}
\renewcommand{\thefigure}{\thesection.\arabic{figure}}
\renewcommand{\theequation}{\Alph{section}.\arabic{equation}}
\section{Obtaining average-power constraint from power profile}
\setcounter{equation}{0}
\label{app:PP}
In Section~\ref{sec:times}, we choose the constraint, given by the time-average power, to be~$\overline{P}=300\,\rm W$.
In this appendix, we show a method of estimating $\overline{P}$ using a power profile, which quantifies a cyclist's power-duration relationship, and is generated from power-meter measurements collected over a period of a ride, a week, and a season.
\begin{figure}
	\centering
	\includegraphics[width=0.65\textwidth]{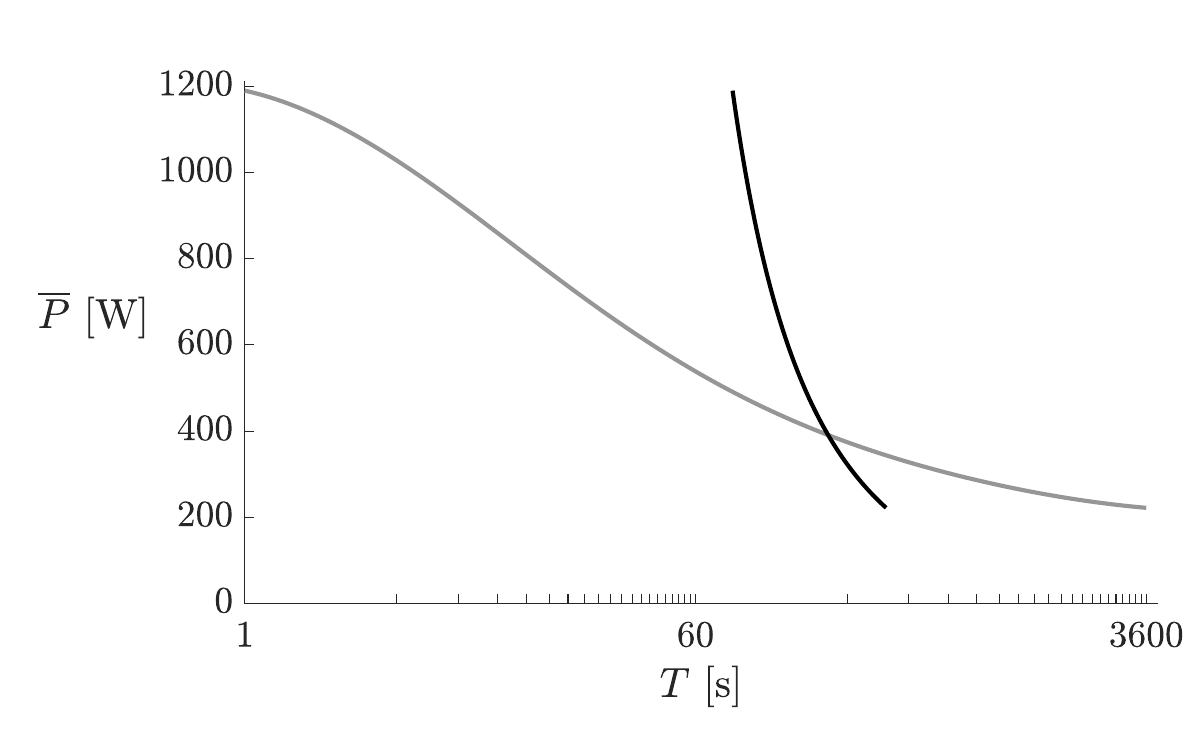}
	\caption{Cyclist's power profile, in grey, and $(T,\,\overline{P})$ pairs for minimum ascent times with power-average constraints,~$200\,{\rm W}\leqslant\overline{P}\leqslant 1200\,{\rm W}$; a logarithmic scale is used on the time axis.}
	\label{fig:Fig_BSSS_Ascent_PP}
\end{figure}

The grey curve in Figure~\ref{fig:Fig_BSSS_Ascent_PP} is a generic power profile \citep[e.g.,][Figure~2]{LeoEtAl2022}.
The black curve is composed of $(T,\,\overline{P})$ pairs calculated for ascent~(\ref{fig:ascenta}), with power-average constraints of $200\,{\rm W}\leqslant\overline{P}\leqslant 1200\,{\rm W}$ and $m=70\,{\rm kg}$, $g = 9.81\,{\rm m/s^2}$, $\rho=1.2\,{\rm kg/m^3}$, ${\rm C_dA} = 0.3\,{\rm m^2}$, ${\rm C_{rr} }= 0.005$, $\lambda = 0.02$, by the method introduced in Section~\ref{sec:Shortest}.

For each power constraint,~$\overline{P}$, we calculate the ground speed,~$V$, and ascent time,~$T$, using expressions~\eqref{eq:vpq} and~\eqref{eq:T}, respectively. 
Since the grade of ascent~(\ref{fig:ascenta}) is constant, the coefficients of the depressed cubic are the same for the constant-power and constant-speed strategies, which results in the same constant~$V$. 

The highest average-power values\,---\,to be used as the constraint for a given ascent time\,---\,are along the grey curve.
The shortest-ascent times\,---\,corresponding to a given average power\,---\,are along the black line.
Hence, the intersection of these curves corresponds to the best performance possible for a given cyclist.
If, for a given ascent time, $\overline{P}$ is below the grey curve, the rider could expend more power; if it is above, it is beyond the rider's capacity.

The sought intersection is calculated numerically by an iterative process.
Herein, its coordinates are $T=200.2343\,\rm s$ and $\overline{P} = 390.6689\,\rm W$.
In accordance with expressions~\eqref{eq:vpq} and~\eqref{eq:VAM}, these values correspond to~$V=5.0191\,\rm m/s$ and~${\rm VAM} = 1797.9005\,\rm m/h$, respectively. 
\section{Power estimate for changes in kinetic energy }
\setcounter{equation}{0}
\setcounter{figure}{0}
\label{app:KE}
As stated in Section~\ref{sub:Formulation}, in this appendix we wish to gain an insight\,---\,albeit indirect and {\it a~posteriori}\,---\,into the justification of neglecting the power associated with changes in kinetic energy.
The insight is indirect because to estimate changes in kinetic energy we consider constant-power, not constant-speed cases, which are the sought solution; this solution necessarily results in $P_K\equiv 0$\,.
It is {\it a~posteriori\/} because we use the formulation wherein changes in kinetic energy are neglected {\it a~priori\/}.
Yet, we claim that our reasoning is not circular; it remains a reasonable estimate of the error committed.

To gain an insight into the power expended to increase kinetic energy, we use the fact that\,---\,for conservative forces\,---\,we need to consider only the initial and final states.
As we see in the right-hand side column of Figure~\ref{fig:PwrSpd}\,---\,and considering the cases of constant power\,---\,the same speed is at the bottom and at the top of ascents~(\subref{fig:ascenta}), (\subref{fig:ascentd}), (\subref{fig:ascente}) of Figure~\ref{fig:ascent}.
As shown in the right-hand side column of Figure~\ref{fig:ascent}, the same speed\,---\,for constant power\,---\,is tantamount to the same slope.
For these ascents, the change of kinetic energy is zero, since
\begin{equation*}
\Delta K = \dfrac{m}{2}\left(V_N^2-V_0^2\right) = 0\,{\rm J},
\end{equation*}
as expected.

Let us consider ascent~(\subref{fig:ascentc}) in Figure~\ref{fig:ascent}, which is steeper at the bottom than at the top. 
In the constant-power case, the speed at the bottom is $V_0=2.1175\,{\rm m/s}$ and the speed at the top is $V_N=11.2361\,{\rm m/s}$.
Hence, the change in kinetic energy is 
\begin{equation*}
	\Delta K = \dfrac{m}{2}\left(V_N^2-V_0^2\right) = 4261.8467\,{\rm J}
\end{equation*}
and\,---\,given $T=263.7539\,{\rm s}$, in Table~\ref{tab:ascents}\,---\,the power required for this increase is
\begin{equation}
\label{eq:PK}
	P_K = \dfrac{1}{1-\lambda}\dfrac{\Delta K}{T} = 16.4882\,{\rm W}.
\end{equation}
Thus, $P_K$ is an order of magnitude less than the specified average power of~$\overline{P}=300\,{\rm W}$.

Examining expression~(\ref{eq:PK}), we see that $P_K$ is inversely proportional to the ascent time.
Hence, neglecting changes in kinetic energy becomes more justifiable for longer ascents.

To quantify this statement, let us return to ascent~(\subref{fig:ascentc}) and scale the horizontal and vertical distances by $1/2$; in other words, we specify $x_n = 500\,{\rm m}$ and $y_n = 50\,{\rm m}$, which maintains the overall grade percentage of $\overline{\mathcal{G}}=10\%$.
Also, the speeds at the bottom and top do not change.
However, the ascent time is halved to $T=131.8770\,{\rm s}$ and, hence, the $P_K$ is doubled to $P_K = 32.9764\,{\rm W}$. 
If we double the horizontal and vertical distances, specifying $x_n = 2000\,{\rm m}$ and $y_n = 200\,{\rm m}$, which also maintains the overall grade percentage, the speeds at the bottom and top do not change but the ascent time is doubled to $T=527.5078\,{\rm s}$ and, hence, the $P_K$ is halved to $P_K = 8.2441\,{\rm W}$, which is two orders of magnitude less than the specified average power.

\section{Average-power constraint subject to maximum power}
\setcounter{equation}{0}
\setcounter{figure}{0}
\label{app:Pmax}
As stated in Section~\ref{sec:Minimum}, a constant speed is a global minimum of the ascent time, if the only constraint is the average power.
However, this solution can require power that might exceed an athlete's capacity. 
In this appendix, we complement the average-power constraint,~$\overline{P}=P_0$, with a maximum power constraint $P_j \leqslant P_{\rm max}$, for $ j = 1,\dots, N$.

Let us call a segment extremal if, in the original problem with only the average power constraint, $P_j > P_{\rm max}$. 
If so, then for that segment we set $P_j = P_{\rm max}$.
As such, the region over which we minimize the ascent time is the space of segment speeds for which $P_j \leqslant P_{\rm max}$\,; consequently, the segments with $P_j > P_{\rm max}$ lie outside of the minimization region.
With the additional constraint of maximum power, the subset of extremal segments becomes
\begin{equation}
	\label{eq:extremalSubset}
	\mathcal{J}
	=
	\{\,j\,\,|\,\, P_j=P_{\rm max}\}
	,\qquad j=1,\dots,N.
\end{equation}
In other words, on extremal segments, the speeds,~$V_\mathcal{J}$, are fixed; their magnitudes are determined using equation~\eqref{eq:Vj_constP}, in the same manner as the constant-power case, except that $P_0$ is replaced with $P_{\rm max}$\,.
Consequently, the time to traverse, as well as the total work done over, all extremal segments is fixed; the former is
\begin{equation}
	\label{eq:TPmax}
    T_{P_{\rm max}} = \sum\limits_{\substack{j=1\\j\in\mathcal{J}}}^N\dfrac{L_j}{V_j},
\end{equation}
and the latter is
\begin{equation}
	\label{eq:WPmax}
    W_{P_{\rm max}} = \sum\limits_{\substack{j=1\\j\in\mathcal{J}}}^NP_{\rm max}\,\dfrac{L_j}{V_j} = P_{\rm max}\,T_{P_{\rm max}}.
\end{equation}

With time~\eqref{eq:TPmax} and work~\eqref{eq:WPmax}, expression~\eqref{eq:Pave} becomes
\begin{equation}
	\label{eq:Pavg_Pmax}
	\overline{P}
	=
	\frac{W}{T}
	=
	\left(\sum\limits_{\substack{j=1\\j\not\in\mathcal{J}}}^N\dfrac{L_j}{V_j} + T_{P_{\rm max}}\right)^{-1}\left(\sum\limits_{\substack{j=1\\j\not\in\mathcal{J}}}^NP_j\,\dfrac{L_j}{V_j} + W_{P_{\rm max}}\right),
\end{equation}
which is the corresponding time-average power for the entire ascent.
For the submaximal segments, the speeds are determined by constraint~\eqref{eq:PaveP0} and, hence, have to be constant.
Thus, following the procedure presented in Section~\ref{sec:speed}, we rewrite the power constraint\,---\,using $V_j=V$ for $j=1,\dots,N$ such that $j\not\in\mathcal{J}$\,---\,as
\begin{equation*}
	0 = V^3 + p\,V + q\,,
\end{equation*}
which is equation~(\ref{eq:DepCub}) with
\begin{equation}
	\label{eq:pq_vconst_Pmax}
	p = \left(\beta\sum\limits_{\substack{j=1\\j\not\in\mathcal{J}}}^NL_j\right)^{-1}\left(\sum\limits_{\substack{j=1\\j\not\in\mathcal{J}}}^N\alpha_jL_j + W_{P_{\rm max}} + P_0\,T_{P_{\rm max}}\right)
\end{equation}
and $q$ the same as in expression~(\ref{eq:pq_vconst}).

Let us revisit ascent~(\ref{fig:ascentd}), for which we calculate ascent times in Section~\ref{sec:times} under the constraint of $\overline{P}=300\,\rm W$ and using a discretization of $N=10\,000$ straight-line segments. 
As stated in Table~\ref{tab:ascents}, the ascent time of $T=254.8642\,\rm s$ is achieved with a constant speed of $V=3.9497\,\rm m/s$ and a maximum power of $\max\{P_j\}=567.6278\,\rm W$. 

To gain insight to the effect of $P_{\rm max}$, let us restrict the power to $P_{\rm max}=400\,\rm W$.
In accordance with expression~\eqref{eq:Pavg_Pmax}, there are $3932$ extremal segments, along which $T_{P_{\rm max}} = 116.3710\,\rm s$ and $W_{P_{\rm max}} = 46548.4088\,\rm J$. 
For the submaximal segments, the constant speed is $V=4.3758\,\rm m/s$, which results in the ascent time of~$T=255.3814\,\rm s$.
\begin{figure}
	\centering
	\includegraphics[width=0.49\textwidth]{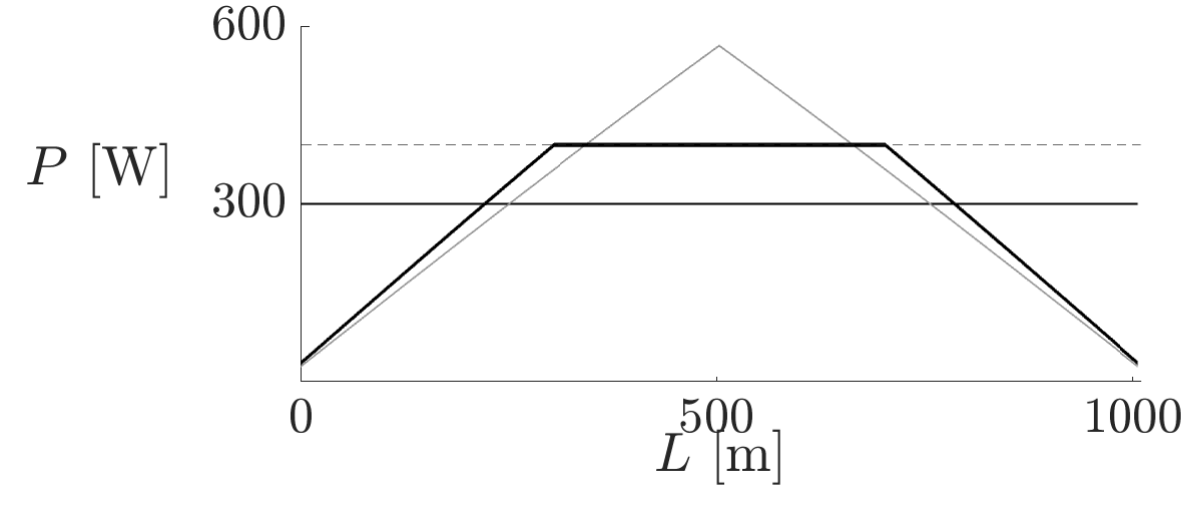}
	\includegraphics[width=0.49\textwidth]{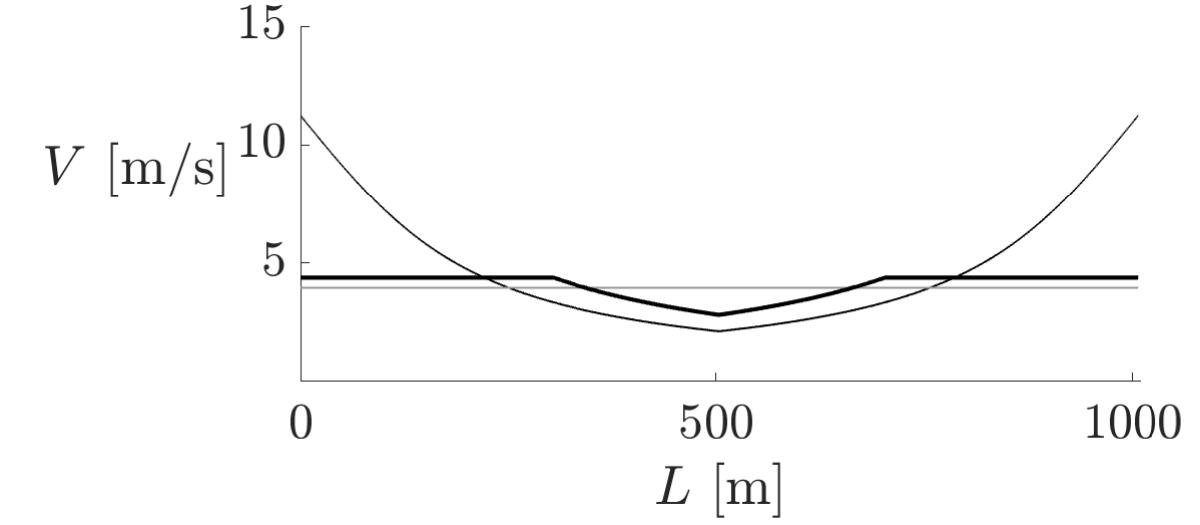}
	\caption{Power and speed corresponding to ascent~(\ref{fig:ascentd}) subject to $P_{\rm max}=400\,\rm W$.
	Thin black lines correspond to the constant-power solution, thin grey lines to the constant-speed solution, thick black lines to the solution subject to $P_{\rm max}$.}
    \label{fig:PwrSpdPmax}
\end{figure}
In this example, which we illustrate in Figure~\ref{fig:PwrSpdPmax}, the effect of $P_{\rm max}$ is to increase the ascent time due to extremal segments. 
If $P_{\rm max} > \max\{P_j\}$, there are no extremal segments and the ascent time is the same as that of the constant-speed solution. 
If $P_{\rm max}\to\overline{P}$, all segments become extremal and the ascent time is the same as for the constant-power solution.
Thus\,---\,even with the $P_{\rm max}$ constraint\,---\,the ascent time is the shortest with, albeit restricted, constant-speed strategy.

Examining Figures~(\ref{fig:ascentd}) and (\ref{fig:PVd}), we see that the power along the less steep parts of the ascent is much lower than one might expect.
However, constraining the power from below on such segments lowers the work budget for the remaining segments and results in an increased time, which is an intrinsic consequence of the average-power constraint.
\section{Numerical optimization considerations}
\setcounter{equation}{0}
\setcounter{figure}{0}
\label{app:NumCon}
Throughout this paper, we use numerical methods to obtain segment speeds that minimize the ascent time subject to nonlinear constraints. 
These speeds can be obtained numerically using any constrained numerical optimization program, such as provided by Matlab. 
However, this approach might be inefficient for ascent discretizations with large~$N$. 
As such, the purpose of this appendix is to demonstrate the efficiency of our formulations, as opposed to standard numerical procedures, for the problem at hand.

Let us turn our attention to the numerical example of Appendix~\ref{app:Pmax}, which pertains to ascent~(\ref{fig:ascentd}) using a discretization of $N=10\,000$ straight-line segments. 
For the solution to the minimum ascent time, where the only constraint is average power $\overline{P}=P_0=300\,\rm W$, we use Matlab's \texttt{fmincon} function~\citep{MATLAB}, which finds the minimum of a constrained nonlinear multivariable function. 
Starting with segment speeds sampled from the uniform distribution, $V_j\sim U(1\,{\rm m/s},10\,{\rm m/s})$, \texttt{fmincon} obtains the solution with~$237$ iterations, requiring~$2\,380\,830$ function evaluations within the optimization algorithm.
By contrast, we obtain the global minimum of the ascent time with a direct calculation using coefficients~\eqref{eq:pq_vconst} in solution~\eqref{eq:vpq}. 

For the solution to the restricted-power problem, we include the $P_{\rm max}=400\,\rm W$ restriction.
Starting with the constant segment speeds that minimize the previous problem, \texttt{fmincon} obtains the solution with $870$ iterations, requiring $8\,711\,250$ function evaluations within the optimization algorithm.
As an alternative, we propose the following iterative procedure that obtains\,---\,for the problem at hand\,---\,the same solution, but typically with fewer than five iterations.

\begin{tabular}{@{}p{0.025\textwidth}p{0.935\textwidth}}
    {\bf P1.} &
    {\rm[Initialize.]} 
    Set segment speeds to the constant speed, $V_j=V$, that corresponds to the global minimum of the ascent time, where the only constraint is the average power, using coefficients~\eqref{eq:pq_vconst} in solution~\eqref{eq:vpq}, set $\mathcal{J}=\emptyset$, and initialize a counter $i=1$.
    \\
    {\bf P2.} &
	{\rm[Check extremal segments.]}
	If all segment powers are $P_j\leqslant P_{\rm max}$, terminate procedure successfully. 
	Otherwise, identify the segments that exceed $P_{\rm max}$ and fix $V_j$ such that $P_j=P_{\rm max}$, using coefficients $p_j=\alpha_j/\beta$ and $q=-P_{\rm max}/\beta$ in solution~\eqref{eq:vpq}; assign these segments as extremal and concatenate to subset $\mathcal{J}$.
	\\
    {\bf P3.} &
	{\rm[Update constant speed.]}
	Recalculate $V_j=V$ for $j\not\in\mathcal{J}$, using~$p$ from expression~\eqref{eq:pq_vconst_Pmax} and~$q$ from expression~\eqref{eq:pq_vconst}, iterate the counter $i=i+1$, and return to P2.
\end{tabular}

Using this procedure, in contrast to \texttt{fmincon}, we obtain the same solution within four iterations only.
The program terminates with $3932$ extremal segments: on the first iteration, $3162$ segments are identified as extremal; on the second, $726$ become extremal due to the updated constant speed; on the third, the remaining $44$ become extremal.

To summarize the procedure, on the first iteration, the segment speeds are set to the value that minimizes the ascent time, under the average-power constraint only. 
In such a case, $3\,162$ segments exceed $P_{\rm max}=400\,\rm W$. 
For these segments, which are extremal, we fix the speeds such that they correspond to $P_{\rm max}$.
For the remaining $10\,000-3\,162=6\,838$ segments, we recalculate the speeds subject to the maximum-power constraint, which is the end of the first iteration.

On the second iteration, we find that $726$ of the $6\,838$ segments exceed $P_{\rm max}$.
We fix the speeds such that they correspond to $P_{\rm max}$ and recalculate the segment speeds for the remaining $6\,838-726=6\,112$ segments, which is the end of the second iteration.

On the third iteration, we find that $44$ of the $6\,112$ segments exceed $P_{\rm max}$.
We repeat the fixing and recalculating of segment speeds, which is the end of the third iteration.

On the fourth iteration, we find that all of the $6\,112-44=6\,068$ segments comply with $P_j\leqslant P_{\rm max}$.
Thus, in the end, the program terminates with $3162+726+44=3932$ extremal segments.

This procedure always terminates in, at most, $N$ iterations as the number of extremal segments is nondecreasing and there are only $N$ segments in total.
In our experience, the procedure converges, in practice, after at most five iterations, but we have no proof of this.
We also cannot guarantee that the computed solution is indeed the global constrained minimum subject to both the average and maximum power constraints, although it seems to be for all the examples we have experimented with.
A proof of this would be interesting. 
\section{Constant ascent speed}
\setcounter{equation}{0}
\setcounter{figure}{0}
\label{app:VertSpeed}
Let us discuss yet another strategy, namely, a constant ascent speed,
for which the vertical component of the ground speed, $V_j\sin\theta_j =: v$, is constant, for $j=1,\dots,N$.
Together with ground speed and power, the ascent speed is provided by a cyclocomputer attached to the handlebars, and hence is a common performance metric.
Following the procedure presented in Section~\ref{sec:speed}, we rewrite the power constraint as $0 = v^3 + p\,v + q$, which is analogous to equation~(\ref{eq:DepCub}), with
\begin{equation}
	\label{eq:pq_vconst_2}
	p 
	=
	\frac{1}{\beta}\dfrac{\sum\limits_{j=1}^N\alpha_jL_j}{\sum\limits_{j=1}^NL_j\csc^2\theta_j}
	\quad\text{and}\quad
	q
	=
	-
	\frac{P_0}{\beta}\dfrac{\sum\limits_{j=1}^NL_j\sin\theta_j}{\sum\limits_{j=1}^NL_j\csc^2\theta_j}\,.
\end{equation}
However, these coefficients present a problem if $\theta_j=0$ and, hence, $\csc^2\theta_j\to\infty$; coefficients~\eqref{eq:pq_vconst_2} tend to zero.
In such cases, the problem of a constant ascent speed is ill-posed.
The cubic equation becomes $v^3=0$\,; hence, $v=0$.
If there is a point where $\theta=0$\,---\,such as the initial or final point in ascents~(\ref{fig:ascentb}) and (\ref{fig:ascentc}), respectively, or the inflection point, in ascents~(\ref{fig:ascentd}) and (\ref{fig:ascente})\,---\,we have an asymptotic ill-posedeness.
If there is a horizontal segment, we have a direct ill-posedeness.

The question of ascent time could be examined further.
The series for $T$ includes terms of the form $v/\sin\theta_j$, which lead to indeterminate forms of~$0/0$.
However, we choose not to engage in elucidating this ill-posed problem.

Be that as it may\,---\,even if ill-posedeness can be avoided by excluding points or segments where $\theta=0$\,---\,a constant ascent speed is not a viable strategy.
For ascent~(\ref{fig:ascenta}), since the grade is constant, the ground speed,~$V=3.9439\,\rm m/s$, power,~$P=300\,\rm W$, and ascent speed,~$v=0.3924\,\rm m/s$, are constant, and result in the same ascent time,~$T=254.8206\,\rm s$\,, as expected in view of Table~\ref{tab:ascents} and $v=V\,\sin(5.7106^\circ)$\,.
In general, however, the instantaneous power and ground speed corresponding to a constant ascent speed are excessively large along less steep portions of the climb in order to keep the vertical component of speed constant.
Notably, this supports the choice of a consistently steep slope to maximize VAM.
\section{On empirical adequacy}
\setcounter{equation}{0}
\setcounter{figure}{0}
\label{app:EmpAdeq}
The empirical adequacy of a phenomenological model needs to be examined by a comparison with measurements.
For this process, we use the ascent\,---\,referred to as the San Bernardo segment\,---\,whose elevation gain is $162\,\rm m$.
As shown in Figure~\ref{fig:SBA}, the slope is variable; the average and maximum grades are $4.8\,\%$ and $8.4\,\%$, respectively.
A recent ascent\,---\,performed specifically for our study\,---\,resulted in $T=543\,\rm s$, which is $9$~minutes and $3$~seconds.
\begin{figure}[h]
	\centering
	\includegraphics[width=0.6\textwidth]{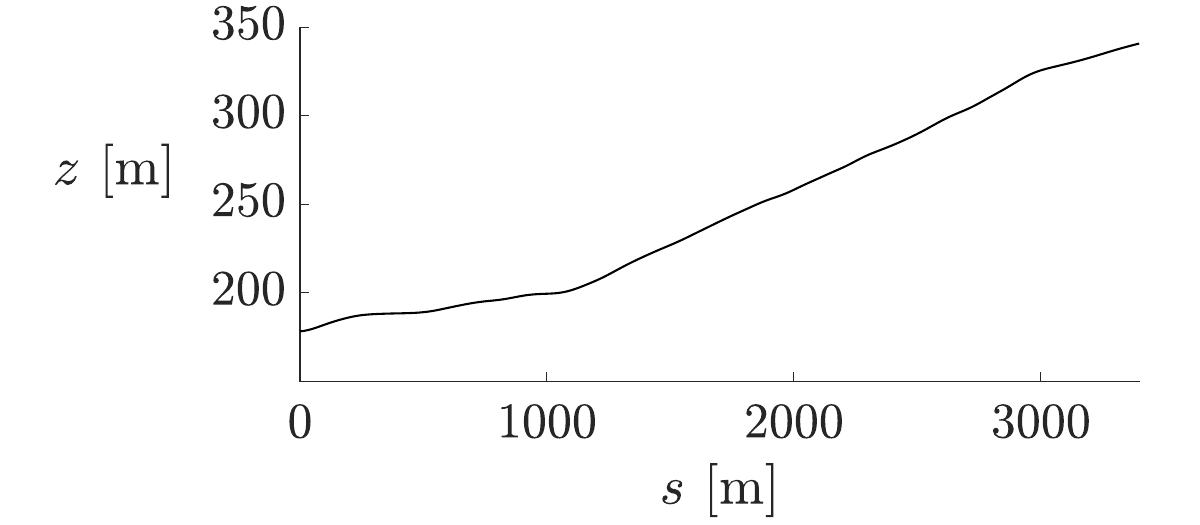}
	\caption{San Bernardo segment, which is a $3$rd category climb in the Italian foothills near Ovada; since the segment has three-dimensional coordinates, the horizontal axis is with respect to arclength,~$s$.}
	\label{fig:SBA}
\end{figure}

\begin{figure}[h]
	\centering
	\includegraphics[width=0.6\textwidth]{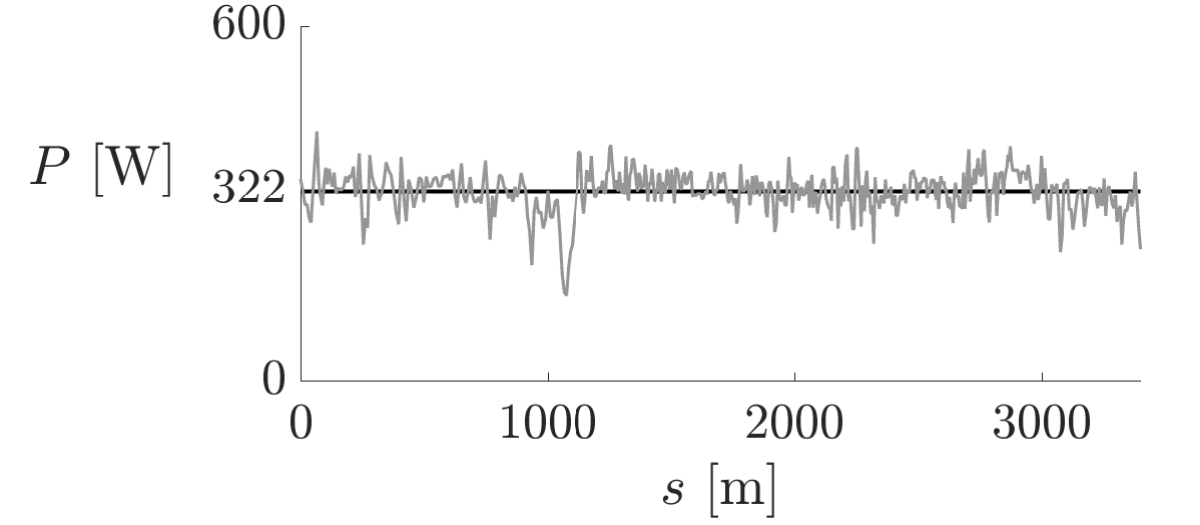}
	\caption{Comparison between average (black) and instantaneous (grey) power}
	\label{fig:SBP}
\end{figure}

The ascent was attempted with constant power, which is easier to maintain than constant speed, since it corresponds to a steady effort.
As shown in Figure~\ref{fig:SBP}, the measured power oscillates about its average of $\overline{P}=322.0699$, in such a manner that its standard deviation is $33.9902\,\rm W$.
These oscillations are\,---\,at least in part\,---\,a consequence of cyclocomputer measurements which are sampled only once a second and, hence, occur necessarily at random instants of pedal revolutions.

To examine the model, we consider the following values in expression~(\ref{eq:PV})\,: $m = 78.6\,\rm kg$, as reported by the rider, ${\rm C_dA}=0.35\,{\rm m}^2$, ${\rm C_{rr}}=0.005$, $\lambda=0.02$, which are common values in this context,%
\footnote{Specific estimates of these three model parameters, for another cyclist, are presented in Appendix~\ref{app:Inverse}.}
$\rho=1.1464\,\rm kg/m^3$, as calculated based on altitude and temperature, and $g=9.81\,\rm m/s^2$.
According to the constant-power strategy\,---\,with $P_0=322\,\rm W$\,---\,$T= 541.7489\,\rm s$, which is $9$~minutes and $2$~seconds; the ascent times agree to within one second.
This specific agreement is obtained by adjusting\,---\,within the range of values expected for the rider in question\,---\,the three model parameters, ${\rm C_dA}$, ${\rm C_{rr}}$ and $\lambda$.
This limited range  of adjustments ensures that the measurements can refute the conjectured model, in the spirit of Popper's conjectures and refutations.
In other words, an agreement between predictions and measurements supports the empirical adequacy of the model.

\begin{figure}[h]
	\centering
	\includegraphics[width=0.55\textwidth]{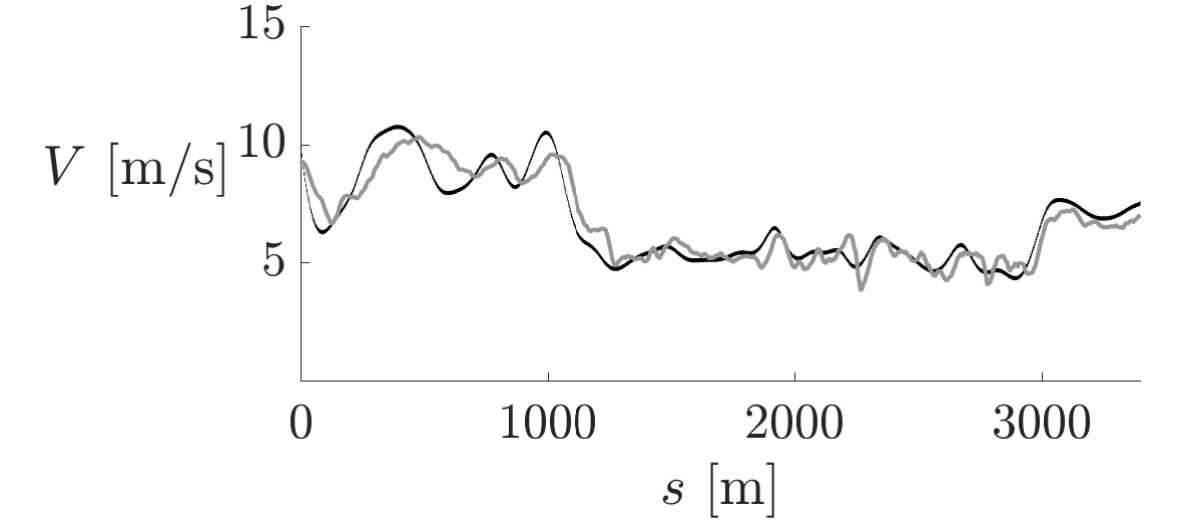}
	\caption{Comparison between modelled (black) and measured (grey) instantaneous ground speeds}
	\label{fig:SBV}
\end{figure}

Another agreement between the model and measurements is the instantaneous speed, shown in Figure~\ref{fig:SBV}.
This agreement is symptomatic of the model's pertinence in examining certain subtleties, not only its capacity of predicting average or global quantities.

This initial examination supports the empirical adequacy of the model.
More rides are planned in the foreseeable future to be followed by a statistical analysis.
Notably, in accordance with our formulation, a constant-speed ascent\,---\,with $\overline{P}=322\,\rm W$\,---\,should result in $T=527.4674\,\rm s$, which is $ 8$~minutes and $47$~seconds, a quarter of a minute less than a constant-power ascent.
\section{Calculus of variations approach}
\label{app:E-L}
\setcounter{equation}{0}
\setcounter{figure}{0}
\label{app:EL}
The discrete model presented in the main text has its continuous counterpart, wherein the minimum ascent time is achieved in the context of the calculus of variations.

Setting $V(s)=\left.{\rm d}s/{\rm d}t\,\right|_{t=t(s)}$, the Lagrangian function stated in expression~\eqref{eq:Lag} is
\begin{equation*}
	\mathcal{L}\left(V,\Lambda\right)=\int\limits_0^T\frac{{\rm d}s}{V} + \Lambda\left(\frac{1}{T}\int\limits_0^T\frac{P}{V}\,{\rm d}s-P_0\right).
\end{equation*}
Multiplying both sides by $T$, we get
\begin{equation*}
	T\,\mathcal{L}=T^2 + \Lambda\left(\int\limits_0^T\frac{P}{V}\,{\rm d}s-T\,P_0\right).
\end{equation*}
Differentiating both sides, we obtain
\begin{equation*}
\left(\nabla T\right)\mathcal{L}+T\left(\nabla\mathcal{L}\right)=2T\nabla T+\Lambda\left(-\int\limits_0^L\frac{P}{V^2}\nabla V\,{\rm d}s+\int\limits_0^L\frac{{\rm d}P}{{\rm d}V}\frac{\nabla V}{V}\,{\rm d}s-P_0\nabla T\right).
\end{equation*}
To obtain a minimum, we set $\nabla\mathcal{L}=0$, for which $\left.\mathcal{L}\right|_{\nabla\mathcal{L}=0}=T$; hence, 
\begin{equation*}
	0=\int\limits_0^L\left((T-\Lambda P_0)\left(-\frac{1}{V^2}\right)-\Lambda\left(\frac{P}{V^2}-\frac{1}{V}\frac{{\rm d}P}{{\rm d}V}\right)\right)\nabla V\,{\rm d}s,
\end{equation*}
which is a form of the Euler-Lagrange equation.
To satisfy this equation for arbitrary integration limits, we require
\begin{equation*}
	0=(T-\Lambda P_0)\left(-\frac{1}{V^2}\right)-\Lambda\left(\frac{P}{V^2}-\frac{1}{V}\frac{{\rm d}P}{{\rm d}V}\right),
\end{equation*}
which we rearrange to write
\begin{equation}
\label{eq:ELeqn}
	0 = \Lambda P_0-T - \Lambda\left(P-V\frac{{\rm d}P}{{\rm d}V}\right)\,.
\end{equation}

To state the expression in parentheses within the context of our model, we invoke equation~(\ref{eq:PV}), namely,
\begin{equation*}
	P = \dfrac{1}{1-\lambda}\left(m\dfrac{{\rm d}V}{{\rm d}t}+mg\left(\sin\theta+{\rm C_{rr}}\cos\theta\right)+\frac{1}{2}{\rm C_dA}\,\rho\,V^2\right)V.
\end{equation*}
Also,
\begin{equation*}
	\frac{\rm d}{{\rm d}V}\left(V\frac{{\rm d}V}{{\rm d}t}\right)
	=
	\frac{{\rm d}V}{{\rm d}V}\frac{{\rm d}V}{{\rm d}t}+V\frac{{\rm d}}{{\rm d}V}\left(\frac{{\rm d}V}{{\rm d}t}\right) 
	=
	(1)\frac{{\rm d}V}{{\rm d}t}+V\left(\frac{{\rm d}}{{\rm d}t}\left(\frac{{\rm d}V}{{\rm d}V}\right)\right)
	=
	\frac{{\rm d}V}{{\rm d}t}+V\left(\frac{{\rm d}}{{\rm d}t}\left(1\right)\right)
	=
	\frac{{\rm d}V}{{\rm d}t}\,.
\end{equation*}
Hence, the derivative in parentheses in expression \eqref{eq:ELeqn} is
\begin{align*}
	\frac{{\rm d}P}{{\rm d}V} 
	&= 
	\frac{\rm d}{{\rm d}V}\left(\dfrac{1}{1-\lambda}\left(mV\dfrac{{\rm d}V}{{\rm d}t}+mg\left(\sin\theta+{\rm C_{rr}}\cos\theta\right)V+\frac{1}{2}{\rm C_dA}\,\rho\,V^3\right)\right)
	\\
	&= 
	\frac{1}{1-\lambda}\left(\dfrac{\rm d}{{\rm d}V}\left(mV\dfrac{{\rm d}V}{{\rm d}t}+mg\left(\sin\theta+{\rm C_{rr}}\cos\theta\right)V+\frac{1}{2}{\rm C_dA}\,\rho\,V^3\right)\right)
	\\
	&= 
	\frac{1}{1-\lambda}\left(m\dfrac{{\rm d}V}{{\rm d}t}+mg\left(\sin\theta+{\rm C_{rr}}\cos\theta\right)+\frac{3}{2}{\rm C_dA}\,\rho\,V^2\right)\,,
\end{align*}
which results in the second term therein,
\begin{equation*}
	V\frac{{\rm d}P}{{\rm d}V}
	=
	\frac{1}{1-\lambda}\left(\left(m\dfrac{{\rm d}V}{{\rm d}t}+mg\left(\sin\theta+{\rm C_{rr}}\cos\theta\right)\right)V+\frac{3}{2}{\rm C_dA}\,\rho\,V^3\right)\,.
\end{equation*}
Subsequently,
\begin{align*}
	\left(1-\lambda\right)\left(P-V\frac{{\rm d}P}{{\rm d}V}\right)
	&=
	\left(\left(m\dfrac{{\rm d}V}{{\rm d}t}+mg\left(\sin\theta+{\rm C_{rr}}\cos\theta\right)\right)V+\frac{1}{2}{\rm C_dA}\,\rho\,V^3\right)
	\\
	&\qquad-
	\left(\left(m\dfrac{{\rm d}V}{{\rm d}t}+mg\left(\sin\theta+{\rm C_{rr}}\cos\theta\right)\right)V+\frac{3}{2}{\rm C_dA}\,\rho\,V^3\right)
	\\
	&=
	-{\rm C_dA}\,\rho\,V^3\,.
\end{align*}
Thus, the Euler-Lagrange equation, stated in expression~(\ref{eq:ELeqn}), becomes
\begin{equation*}
	0 = \Lambda P_0-T-\Lambda\left(-{\rm C_dA}\,\rho\,V^3\right)\,,
\end{equation*}
which implies that
\begin{equation*}
{\rm C_dA}\,\rho\,V^3 = \frac{T}{\Lambda} - P_0\,,
\end{equation*}
and therefore $V$ is a constant.

To find that constant, we must satisfy the average power constraint,
\begin{equation*}
	\frac{1}{T}\int\limits_0^TP\,{\rm d}t=P_0\,.
\end{equation*}
Since ${\rm d}s/{\rm d}t = V\implies{\rm d}t ={\rm d}s/V$,
\begin{equation*}
	\frac{1}{T}\int\limits_0^LP\,\frac{{\rm d}s}{V}=P_0\,.
\end{equation*}
Since for a constant speed, $T=L/V$,
\begin{equation*}
	\frac{V}{L}\int\limits_0^LP\,\frac{{\rm d}s}{V}=\frac{1}{L}\int\limits_0^LP\,{\rm d}s=P_0\,.
\end{equation*}
Returning to equation~(\ref{eq:PV}),
\begin{equation*}
	P = \dfrac{1}{1-\lambda}\left(m\dfrac{{\rm d}V}{{\rm d}t}+mg\left(\sin\theta+{\rm C_{rr}}\cos\theta\right)+\frac{1}{2}{\rm C_dA}\,\rho\,V^2\right)V,
\end{equation*}
with ${\rm d}V/{\rm d}t=0$,  for a constant $V$, we write
\begin{align*}
	\left(1-\lambda\right)\int\limits_0^LP\,{\rm d}s 
	&=
	\int\limits_0^L\left(m\left(0+g\left(\sin\theta+{\rm C_{rr}}\cos\theta\right)\right)V+\frac{1}{2}{\rm C_dA}\,\rho\,V^3\right){\rm d}s 
	\\
	&=
	m\,g\,V\int\limits_0^L\left(\sin\theta+{\rm C_{rr}}\cos\theta\right){\rm d}s+\frac{1}{2}{\rm C_dA}\,\rho\,V^3\int\limits_0^L\left(1\right){\rm d}s
	\\
	&=
	m\,g\,V\int\limits_0^L\left(\sin\theta+{\rm C_{rr}}\cos\theta\right){\rm d}s+\frac{1}{2}{\rm C_dA}\,\rho\,V^3L
\end{align*}

If the ascent profile is given by $\left(x(s),y(s)\right)$, where $0\leqslant s\leqslant L$, and parameterized by arclength, $\left(x'(s),y'(s)\right) = \left(\cos\theta,\sin\theta\right)$, the unit tangent vector is
\begin{equation*}
	\cos\theta=x'(s)\quad\text{and}\quad\sin\theta=y'(s).
\end{equation*}
Hence
\begin{equation*}
    \int\limits_0^L\left(\sin\theta+{\rm C_{rr}}\cos\theta\right){\rm d}s
 	=
    \int\limits_0^L\left(y'(s)+{\rm C_{rr}}x'(s)\right){\rm d}s
    =
    \left(y(L)-y(0)\right)+{\rm C_{rr}}\left(x(L)-x(0)\right)
    =
    H+{\rm C_{rr}}\,D,
\end{equation*}
where $H:=y(L)-y(0)$ is the height gain and $D:=x(L)-x(0)$ is the horizontal gain.

It follows that
\begin{equation*}
	\left(1-\lambda\right)\int\limits_0^LP\,{\rm d}s
	=
	m\,g\,V\left(H+{\rm C_{rr}}\,D\right) + \frac{1}{2}{\rm C_dA}\,\rho\,V^3L
\end{equation*}
and thus the average power constraint is
\begin{equation*}
	\frac{1}{L}\int\limits_0^LP\,{\rm d}s = P_0
	\iff
	\frac{m\,g\,V\left(H+{\rm C_{rr}}\,D\right)}{(1-\lambda)L} + \dfrac{\tfrac{1}{2}{\rm C_dA}\,\rho\,V^3}{1-\lambda} = P_0\,.
\end{equation*}
This is a cubic equation for $V$, which can be solved numerically; there is only one positive real solution. 
There is also the somewhat laborious, yet analytical, Cardano's solution.

To exemplify this formulation, let us consider $m = 70\,{\rm kg}$, ${\rm C_dA} = 0.3\,{\rm m^2}$, ${\rm C_{rr}}= 0.005$, $\lambda = 0.02$, $g = 9.81\,{\rm m/s^2}$, $H = 100\,{\rm m}$, $D = 1000\,{\rm m}$,  $L = 1004.9876\,{\rm km}$,  $\rho = 1.2\,{\rm kg/m^3}$ and $P_0 = 300\,{\rm W}$, which results in
\begin{equation*}
	V^3+398.5870 V  = 1633.3333\,.
\end{equation*}
Its only real solution is $V=3.9439\,{\rm m/s}$, as expected in view of the first row in Table~\ref{tab:ascents}.
Letting $L=1006.6272\,{\rm m}$, we obtain a slightly different cubic whose only real solution is the speed for the constant-speed strategy in the remaining rows of that table.
In all cases, the corresponding ascent times are $T=L/V$\,.

The constant-power case leads to a first order nonlinear differential equation for $V(t)$\,; hence, in contrast to the constant-speed case presented herein, it has no explicit solution.
\section{On estimation of resistance parameters}
\label{app:Inverse}
\setcounter{equation}{0}
\setcounter{figure}{0}
\label{app:PrmEst}
In Appendix~\ref{app:EmpAdeq}, we demonstrate the empirical adequacy of power model~\eqref{eq:PV} through its agreements with measurements of ascent time and instantaneous speeds of a cyclist along the San Bernardo ascent. 
These agreements depend on accurate estimates of the air, ${\rm C_dA}$, rolling, ${\rm C_{rr}}$, and drivetrain, $\lambda$, resistance parameters, which we obtain by adjusting their values within the expected range.
In this appendix, we demonstrate a procedure to estimate these values using numerical optimization.

For this purpose, we obtain further measurements along the San Bernardo ascent; these ascents are done by another cyclist.
Following the approach of Appendix~\ref{app:EmpAdeq}, the cyclist used a flying start and the constant-power strategy.
For a statistical evaluation, the cyclist repeated the ascent three times, striving to maintain the same power along a given ascent.

The data are collected using a Garmin~1030 cyclocomputer in combination with pedal-based 4iii power metres.
The data are sampled once a second and measure altitude [m], ambient temperature [${}^\circ{\rm C}$], cadence [rpm], distance [m], heartrate [bpm], power [W], time [s], and velocity [m/s]. 
The environmental conditions during the efforts correspond to a windless and sunny day with a temperature of approximately $22^\circ{\rm C}$. 
Using the temperature, along with the altitude, we estimate the air density,~$\rho=1.1565\,{\rm kg}/{\rm m}^3$, using the formula presented in~\citet[Section~5.3]{DanekEtAl2021}. 
Measurements of ascent time, $\mathcal{T}$, average power, $\overline{\mathcal{P}}$, and distance travelled, $L$, are summarized in Table~\ref{tab:opt} for each effort.
We note that\,---\,in spite of the fact that each ascent corresponds to the same Strava%
\footnote{Strava is an internet service for tracking physical exercise using GPS.}
segment\,---\,the distance travelled is different for each effort, which is due to the rather sparse sampling of the cyclocomputer and the position error of the GPS.
\begin{table}[h]
	\centering
	\begin{tabular}{c*{8}{c}}
		& \multicolumn{3}{c}{Data measurements} & & \multicolumn{3}{c}{Optimized resistance parameters} \\
		\cmidrule{2-4}\cmidrule{6-8}
		effort & $\mathcal{T}\,{\rm[s]}$ & $\overline{\mathcal{P}}\,{\rm[W]}$ & $L\,{\rm[m]}$ & & ${\rm C_dA}\,[{\rm m}^2]$ & ${\rm C_{rr}}$ & $\lambda$ \\
		\toprule
		1 & $579$ & $247\pm13$ & $3486$ & & $0.3521\pm0.0128$ & $0.0059\pm0.0005$ & $0.0317\pm0.0017$ \\
		2 & $577$ & $247\pm15$ & $3426$ & & $0.3671\pm0.0102$ & $0.0062\pm0.0004$ & $0.0321\pm0.0016$ \\
		3 & $489$ & $312\pm27$ & $3467$ & & $0.3575\pm0.0084$ & $0.0059\pm0.0004$ & $0.0313\pm0.0010$ \\
		\bottomrule
	\end{tabular}
	\caption{Optimization of resistance parameters for three efforts of the San Bernardo ascent following the constant-power strategy}
	\label{tab:opt}
\end{table}

To estimate the resistance parameters in model~\eqref{eq:PV}, we require that ${\rm C_dA}$, ${\rm C_{rr}}$, and $\lambda$ result in a modelled ascent time, $T$, that agrees with the measured ascent time, $\mathcal{T}$. 
In other words, we seek to obtain numerical solutions to $f({\boldsymbol\beta}) = \left(T({\boldsymbol\beta}) - \mathcal{T}\right)^2$, where ${\boldsymbol\beta}=\left[\,{\rm C_dA},\,{\rm C_{rr}},\,\lambda\,\right]^T$, such that the power along each segment of an effort is fixed to the measured average power, $\overline{\mathcal{P}}$.
We solve the problem using~\citet{MATLAB}'s~\texttt{fmincon} solver, which finds the minimum of a constrained nonlinear multivariable function. 
To maintain reasonable estimates, we apply bounds to the resistance parameters in the optimization such that ${\rm C_dA}\in[0.1,0.5]\,{\rm m}^2$, ${\rm C_{rr}}\in[0.001,0.01]$, $\lambda\in[0.01,0.05]$. 
Also, in contrast to the cyclist in Appendix~\ref{app:EmpAdeq}, the mass of the bicycle-cyclist system is $m=60.6\,{\rm kg}$. 

Given the underdetermined nature of the problem, there are many combinations of ${\rm C_dA}$, ${\rm C_{rr}}$ and $\lambda$ that result in an agreement between modelled and measured ascent time, $T=\mathcal{T}$, under the assumption of constant power, $P_j=\overline{\mathcal{P}}$\,, for $j=1,\dots,N$.
For this reason, we repeat the optimization $10\,000$ times for each effort using different starting resistance values sampled uniformly from the specified parameter bounds. 
The sample mean and standard deviation of the optimized resistance parameters are stated in Table~\ref{tab:opt}.
We see that the resistance-parameter estimates are similar for each of the three ascents, which is a supporting evidence of empirical adequacy of the model.
\end{appendix}
\end{document}